\documentclass[usegraphicx,usedcolumn,usenatbib]{mn2e}
\usepackage{xspace}
\usepackage{color,epsfig,rotating,amssymb,longtable}
\usepackage{array,colortbl}
\usepackage{lscape}
\usepackage[colorlinks=true,linkcolor=magenta,citecolor=blue,breaklinks=true]{hyperref}
\usepackage{longtable}
\usepackage{ifpdf}

\ifpdf
\usepackage{epstopdf}
\else
\usepackage[dvips]{graphicx}
\usepackage{lscape}
\fi

\newcommand\ion[2]{#1\,{\sc {#2}}\relax} 

\newcommand{\HII}{H{\sc ii}\xspace}

\newcommand{\OIII}{\ion{O}{iii}}
\newcommand{\OII}{\ion{O}{ii}}
\newcommand{\OI}{\ion{O}{i}}
\newcommand{\SIII}{\ion{S}{iii}}
\newcommand{\SII}{\ion{S}{ii}}
\newcommand{\NII}{\ion{N}{ii}}

\newcommand{\HeI}{\ion{He}{i}}
\newcommand{\HeII}{\ion{He}{ii}}
\newcommand{\NeIII}{\ion{Ne}{iii}}
\newcommand{\FeIII}{\ion{Fe}{iii}}

\newcommand{\ArIII}{\ion{Ar}{iii}}
\newcommand{\ArIV}{\ion{Ar}{iv}}

\def\Haro{Haro\,15\xspace}

\usepackage{color}

\definecolor{dgreen}{rgb}{0,.5,.1} 
\definecolor{pink}{rgb}{.9,.4,.7}

\usepackage{graphicx}
\title[Haro 15: Physical properties]{High resolution spectroscopy of the BCD galaxy Haro\,15: II. Chemodynamics.} 

\author[G.~F. H\"agele et al.]{Guillermo F. H\"agele$^{1,2,3}$,
  Ver\'onica Firpo$^{1,2}$, Guillermo Bosch$^{1,2}$ 
 \newauthor
 \'Angeles I. D\'{\i}az$^{3}$ and Nidia Morrell$^{4}$ \\   
$^{1}$ Facultad de Ciencias Astron\'omicas y Geof\'{\i}sicas, Universidad Nacional  
de la La Plata, Paseo del Bosque s/n, 1900 La Plata, Argentina.\\ 
$^{2}$ IALP-Conicet, Paseo del Bosque s/n, 1900 La Plata, Argentina.\\ 
$^{3}$ Departamento de F\'{\i}sica Te\'orica, C-XI, Universidad Aut\'onoma de
Madrid, 28049 Madrid, Spain.\\
$^{4}$ Las Campanas Observatory, Carnegie Observatories, Casilla 601, La Serena, Chile.} 

\begin{document} 

\date{Accepted . Received ; in original form } 


\maketitle

\begin{abstract} 
We present a detailed study of the physical properties of the nebular material
in four star-forming knots of the blue compact dwarf galaxy \Haro. Using
long-slit and echelle spectroscopy obtained at Las Campanas Observatory, we
study the physical conditions (electron density and temperatures), ionic and
total chemical abundances of several atoms, reddening and ionization
structure, for the global flux and for the different kinematical
components. The latter was derived by comparing the oxygen and sulphur ionic
ratios to their corresponding observed emission line ratios (the $\eta$ and
$\eta$' plots) in different regions of the galaxy. Applying the direct method or
empirical relationships for abundance determination, we perform a comparative
analysis between these regions. The similarities found in the ionization structure of the different kinematical components implies that the effective temperatures of the ionizing radiation fields are very similar in spite of some small differences in the ionization state of the different elements. Therefore the different gaseous kinematical components identified in each star forming knot are probably ionized by the same star cluster. However, the difference in the ionizing structure of the two knots with knot A showing a lower effective temperature than knot B, suggests a different evolutionary stage for them consistent with the presence of an older and more evolved stellar population in the first.
\end{abstract}


\section{Introduction}

Giant Extragalactic H\,{\sc ii} Regions (GH{\sc ii}Rs) are extended,
 luminous objects, which are observed on the discs of spirals and in
 irregular galaxies. GH{\sc ii}Rs are formed due to the presence 
of a large number of young and massive stars whose ultraviolet flux ionizes the 
surrounding gas.
The behaviour of emission lines observed in \HII galaxies and {Blue Compact
Dwarfs (BCDs)} spectra
resembles those of Giant \HII Regions. We can, therefore, use the same
analysis techniques developed for the latter to derive the temperatures, densities
and chemical composition of the interstellar gas in the star formation bursts
detected in these low metallicity galaxies
\citep{1970ApJ...162L.155S,1980ApJ...240...41F,1991A&AS...91..285T}. BCD
galaxies present strong starbursts easily identified through their strong and
narrow emission lines, low metallicity environments and a complex history of
star formation. These facts make them interesting objects to study metallicity
effects in galaxies \citep{2000A&ARv..10....1K}. 

As mentioned in Firpo et al.\ (\citeyear[hereafter Paper I]{Firpo+11}) \Haro
has an absolute magnitude M$_{B}$ = -20.69, a surface brightness $ \mu $=
18.56 mag arcsec$ ^{-2} $ and a colour B-V= 0.33
\citep{2001ApJS..133..321C}. At a distance of 86.6 Mpc \citep{RC3.9}, Haro 15
meets the criteria for a Luminous Compact Blue Galaxy (LCBG)
\citep{2004AJ....128.1541H}, although some authors suggest that it may represent the final outcome of a merger between a dwarf elliptical and a gas
rich dwarf galaxy or HI cloud \citep{Ostlin98,2008A&A...479..725C}. The
interactions taking place during the merging process would act as the
starburst trigger.

\cite{2005A&A...437..849S} derived an overall average oxygen abundance of 12+log(O/H)\,=\,8.33 for the Haro\,15 galaxy, based on an {electron} temperature determination of
8330\,K. It was, however, previously known from H$\alpha$ imaging
\citep{2001ApJS..133..321C} that the galaxy shows a complex morphology, where
the star forming region can be split among a large number of knots scattered
throughout the whole galaxy. \cite{2009A&A...508..615L} obtained spectroscopic
information 
from individual regions and determined densities, temperatures and chemical
abundances for several knots within Haro\,15. Oxygen abundances were derived
for regions A and B (names following  Cair\'os et
al.,~\citeyear{2001ApJS..133..321C}) and found to be 8.37$\pm$0.10 and 8.10$\pm$0.06, respectively. 

\cite{1992ApJ...390..536E} introduced the term ''chemodynamics" after merging
chemical and kinematical information in their analysis of the Wolf-Rayet ring
nebula NGC\,6888. In their research, these authors combined high resolution
spectra with observations at high spatial resolution for different areas of
the nebula, which enabled them to analyse the different velocity components of
the studied region. 

\cite{2009MNRAS.398....2J} studied the BCD galaxy Mrk\,996 using high spectral resolution data obtained with VLT VIMOS in its IFU mode. They found that this galaxy shows multiple components in its emission line profiles, evident as relatively broad and narrow components, and were able to deblend them in order to analyse their physical properties separately. According to the authors this peculiar BCD galaxy has extremely dense gas in its nuclear region, where stellar outflows and shock fronts contribute to generate the relatively broad feature. However James and collaborators could not find evidence of complex emission line profiles for the auroral lines sensitive to temperature. The [\OIII]~$\lambda$\,4363 and [\NII]~$\lambda$\,5755 emission lines, present in their spectra, showed simple profiles which could be fitted
by a broad Gaussian function.
 

In a previous work we presented
a detailed study of the internal kinematics of the nebular material in
multiple knots of the blue compact dwarf galaxy \Haro (Paper I) from echelle spectroscopy. In that paper we
have performed a thorough analysis of its emission lines, including multiple
component fits to the profiles of their strong emission lines. Our
results have shown that the giant \HII\ regions of \Haro\ present a complex
structure in all their emission profiles, detected both in recombination and
forbidden lines. The emission lines of the brightest region, \Haro\,A, can be
split in at least two strong narrow components plus a broad one. Although the
observed regions tend to follow the galaxy kinematics, the components of knot
A have relative velocities that are 
too large to be explained by galactic rotation. Almost all knots follow the
relation between luminosity and velocity dispersion found for virialized
systems, either when considering single profile fittings or the strong narrow
components in more complex fits. The one component fits show a
relatively flatter slope. 


Generally, the flux calibration of echelle data is difficult to check and the
weak auroral emission lines do not have enough signal-to-noise ratio (S/N) to
allow the accurate measurement necessary for a reliable estimate of the
temperature of the emitting gas. Long-slit spectroscopic data of intermediate
resolution are more sensitive and more reliably calibrated to perform temperature estimates and abundance analysis. The combination of echelle
and long-slit data provide an opportunity to interpret the abundance results in
the light of the internal kinematics of the region while providing important
checks on the spectrophotometric calibration. 

The development of Integral Field Unit (IFU)
instruments has provided the spatial
coverage requiered to study extended galactic or extra galactic star-forming regions 
\citep[see e.g.][]{Relano+10,Cairos+10,Monreal-Ibero+10,Monreal-Ibero+11,Rosales-Ortega+10,Rosales-Ortega+11,Perez-Gallego+10,Garcia-Benito+10,Sanchez+10,Lopez-Sanchez+11,Perez-Montero+11}. 
However, slit spectroscopy (with medium or high spectral resolution) is a valid option for
spectrophotometric analysis. This occurs when the object is very compact, or even extended but
with few star-forming knots, and when good spatial and
spectral resolution together with a simultaneous wide spectral coverage are required
\citep[see e.g.][]{Cumming+08,Firpo+10,Firpo+11,2006MNRAS.372..293H,H07,2008MNRAS.383..209H,2008PhDT........35H,2009MNRAS.396.2295H,Hagele+10,2011MNRAS.tmp..430H,Perez-Montero+09,2009A&A...508..615L,2010A&A...516A.104L,Lopez-Sanchez+10b,2010A&A...521A..63L}.

In this paper, we perform a detailed analysis of the physical characteristics of the ionized gas in \Haro, distinguishing among individual components and global line fluxes. {Electron} temperatures and densities were estimated and used to derive ionic and total abundances for all the different species such as: O, S, Ne and Ar.
In what follows, we describe the details of the observations and the relevant data reduction procedures. Section
\ref{sec:rvresults} explains the different methods used to determine the physical conditions of the ionized gas which led to the calculation of their chemical abundances, together with their uncertainties. The results for each region from each applied technique are discussed in Section \ref{sec:Discus}, and the summary and conclusions are presented in Section \ref{secConclus}.


\section{Observations and Data Reduction} 
\label{sec:observations} 

We obtained a long-slit moderate resolution spectrum using the Wide-Field CCD
(WFCCD) camera (September 28 of 2005) mounted on
the 100-inch du Pont Telescope, Las Campanas Observatory (LCO). The 
observations correspond to knots B and C of the BCD galaxy \Haro\ (see Fig.\
1 of Paper I). The TEK5 detector was used covering the wavelength range 3800 to
9300~\AA\ (centred at $\lambda_ {c}$\,=\,6550\,\AA). The effective slit width
was 1\arcsec\ giving a spectral resolution of R\,$\simeq$\,900
($\Delta\lambda$\,=\,7.5\,\AA\ at $\lambda$\,6700\,\AA), as measured
from the {\sc FWHM} of the He-Ne-Ar comparison lines taken for wavelength
calibration purposes. Observing conditions were good,
with an average seeing of 1\arcsec\ and photometric sky conditions.

Bias and dome flat-field frames were taken at the beginning of the night. The
images were processed and analysed with IRAF\footnote{Image Reduction and  
Analysis Facility, distributed by NOAO, operated by AURA, Inc., under
agreement with NSF.} routines following standard procedures. The procedure
includes the removal of cosmic rays, bias-substraction, division by a
normalized flat-field and wavelength-calibration. {To substract the sky
emission we defined two windows at both sides of the 2D spectrum of each
observed knot. For knot B the sky substraction is very good. The final
spectrum shows relatively weak residuals of the OH sky telluric emission lines
which not affect any important emission line from the star-forming region. In
the case of knot C, we have a very good sky substraction for the blue part of
the spectrum ($\lambda\,<$\,~7000\,\AA). We do not measure any emission line
in the red part of this spectrum. The goodness of the sky substractions were
taken into account in the estimated errors of the measurements of the emission
lines. } The standard star EG\,131 was observed for flux calibration purposes
and the exposure time for this flux standard star was 180 seconds. The spectra
were corrected for atmospheric extinction and flux calibrated.

A certain degree of second order contamination is present in our long-slit
spectrum at wavelengths larger than about 6000\,\AA. We are not able to 
deconvolve the different contributions of each order in the red end of the
spectral range, thus, the measured fluxes of the emission lines in that part
of the spectrum may have been slightly overestimated, but these contributions
are not very important for the line ratios.  

Five different regions in \Haro\ (see Fig. 1 of Paper I) were observed with
high resolution spectroscopy obtained using an echelle spectrograph attached
to the 100-inch du Pont Telescope, Las Campanas Observatory (LCO), in July 19
and 20 of 2006. The spectral range of the observations covers from 3400
to 10000\,\AA. This spectral range guarantees the simultaneous measurement of
the nebular emission lines from [\OII]\,$\lambda\lambda$\,3727,3729 to
[\SIII]\,$\lambda\lambda$\,9069,9532\,\AA\ at both ends of the spectrum, in
the very same region of the galaxy. Observing conditions were good with an
average seeing of 1\arcsec\ and photometric nights.  


A detailed description about the echelle data and the reduction procedure can be
found in Paper I. The effective slit width and length were 1\arcsec\ and 4\arcsec\
respectively, and the spectral resolution achieved in our du Pont Echelle data
was R$\simeq$25000 ($\Delta\lambda$=0.25\AA\ at $\lambda$\,6000\AA), as measured
from the FWHM of the Th-Ar comparison lines taken for wavelength calibration
purposes. 
The spectra were obtained as single exposures of 1800 seconds
each. Flux calibration was performed by observing, with an exposure time of
1200\,s, the CALSPEC spectrophotometric standard star Feige\,110
\citep{Bohlin01}. This star flux is tabulated every 2\AA, which is ideal for
calibrating high resolution echelle spectra. In addition, Th-Ar comparison
spectra, milky flats (sky flats obtained with a diffuser, during the
afternoon) and bias frames were taken every night. A journal of observations
is shown in Table \ref{Regions}. The corrected data were reduced with {\sc IRAF}
routines following procedures similar to those described in \cite{Firpo05}.
It must be noted that data obtained for the fainter knot F have not sufficient signal to noise ratio as to allow flux measurements with the accuracy needed to perform an analysis of the physical conditions of the gas and have therefore been excluded from this paper.



\begin{table}
\caption[Observed dates]{Journal of observations for the knots. The first
  column is the observation mode, the second column is the nomenclature used
  in Paper I, the third column  is the observed date, the fourth column is the
  exposure time for the knots, the fifth column is the air masses.} 
\label{Regions}
\begin{center}
\begin{tabular}{@{}ccccc@{}}
\hline 
Mode & knots & Date  & Exp.\ time &sec z  \\
\hline
long-slit&B,C& 2005 Sep 28 & 2x1200\,+\,1x900 & 1.5  \\[2pt]
echelle  & A & 2006 Jul 19 & 1800 & 1.2 \\
         & B & 2006 Jul 19 & 1800 & 1.1 \\  
         & C & 2006 Jul 19 & 1800 & 1.1 \\
         & E & 2006 Jul 20 & 1800 & 1.2 \\
\hline 
\end{tabular}
\end{center}
\end{table}

\section{Results}
\label{sec:rvresults} 

\subsection{Line intensities and reddening correction}

\begin{figure*}
\centering
\includegraphics[angle=0,width=.7\textwidth,height=.4\textwidth]{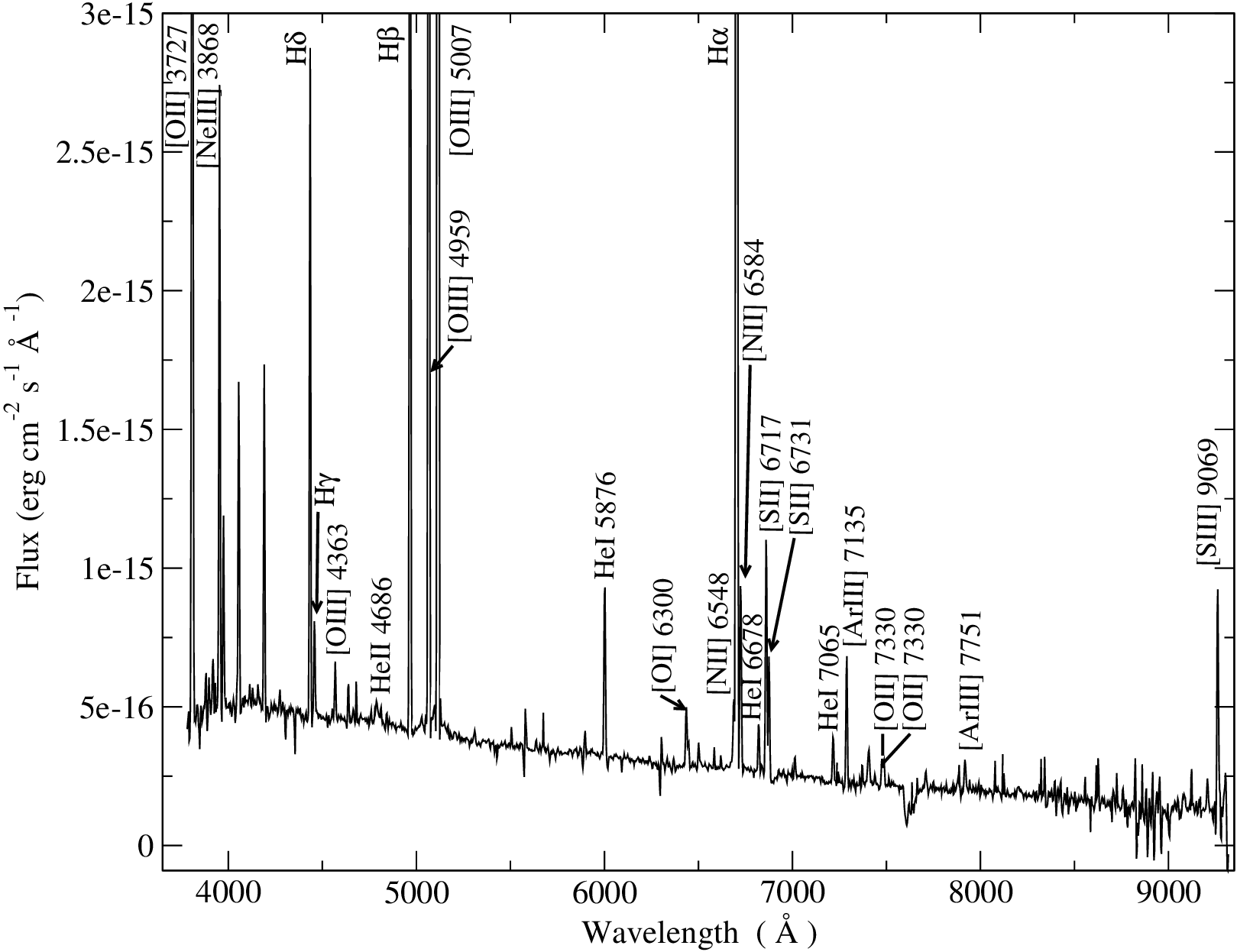}\\
\vspace{0.5cm}
\includegraphics[angle=0,width=.7\textwidth,height=.4\textwidth]{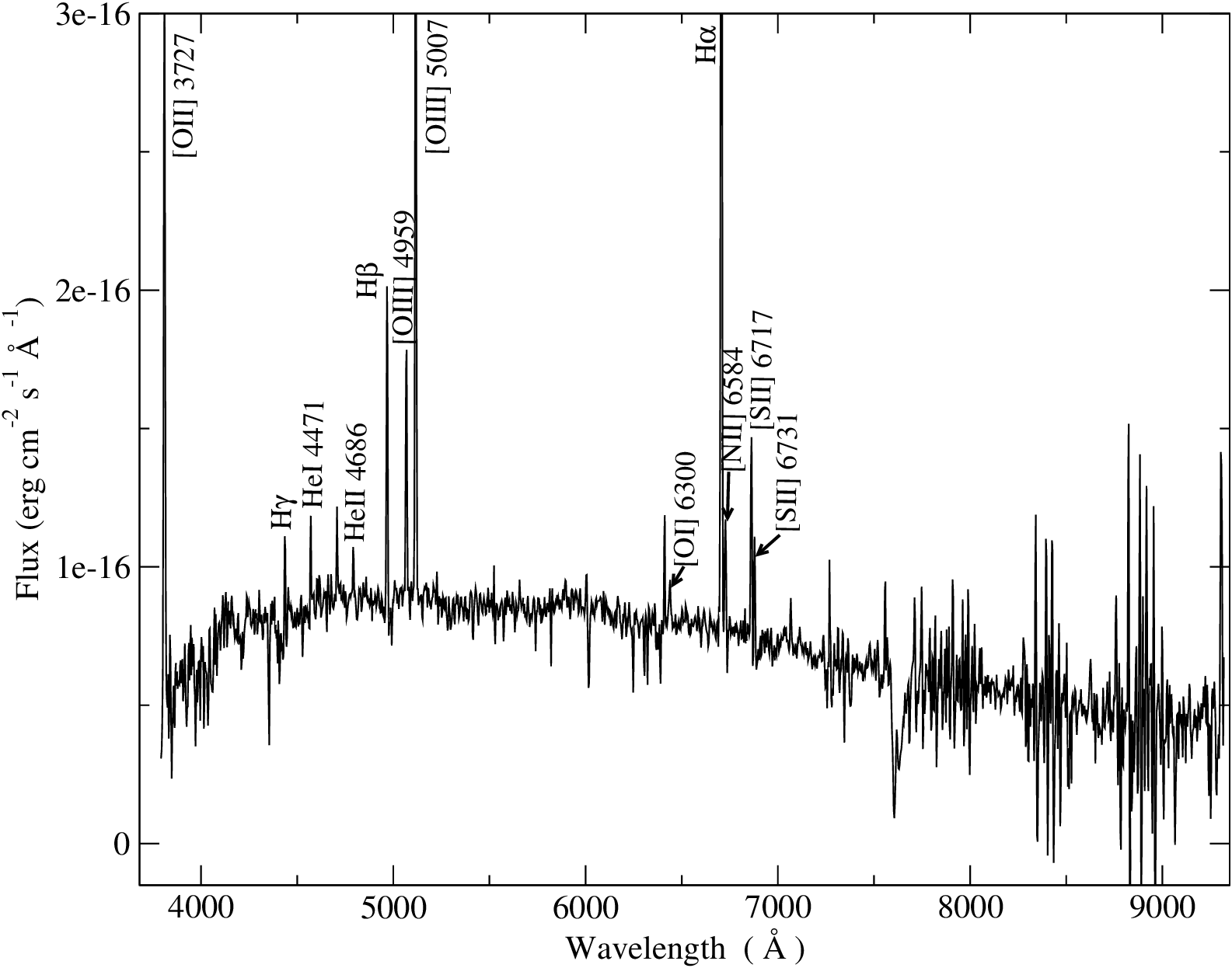}
\caption[BC-spectra]{Long-slit spectra of knots B and C of \Haro\ (upper and lower panel, respectively).}
\label{fig:BCsd}
\end{figure*}

 The one dimensional spectra of knots B and C extracted from the WFCCD
long-slit data are shown in Fig.\ \ref{fig:BCsd}. The spectral plots also include some of the relevant identified emission lines.

For these long-slit data the emission line fluxes were measured using the
\texttt{splot} task in {\sc IRAF} following the procedure described in
\cite{2006MNRAS.372..293H}. We 
used two methods to integrate the emission line flux: (i) In the case of an
isolated line or two blended and unresolved lines, the intensity was
calculated integrating between two points given by the position of the local
continuum placed by eye; (ii) if two lines are blended, but they can be
resolved, we have used a multiple Gaussian fit procedure to estimate
individual fluxes. Both procedures are well detailed in H\"agele et
al.\ (\citeyear[hereafter H06 and
  H08, respectively]{2006MNRAS.372..293H,2008MNRAS.383..209H}). 
In the case of the echelle data we have also followed the procedure described
in (i) to measure the flux of a given line including all the different
kinematical components. This flux will be referred to as ``global" in this work. 

The statistical errors associated with the observed emission fluxes have been
calculated using the expression \[ \sigma_{l}\,=\,\sigma_{c}N^{1/2}[1 +
  EW/(N\Delta)]^{1/2} \] \noindent following the procedure described in
\cite{2003MNRAS.346..105P} where $\sigma_{l}$ is the error in the observed line
flux, $\sigma_{c}$ represents the standard deviation in a box near the
measured emission line and stands for the error in the continuum placement, N
is the number of pixels used in the measurement of the line flux, EW is the
line equivalent width expressed in \AA, and $\Delta$ is the wavelength dispersion in \AA\ per pixel \citep{1994ApJ...437..239G}.  

Following the iterative procedure presented in \cite{Firpo+10}, we measured the
fluxes {and estimated the errors} associated with each kinematical
component identified in the high 
resolution echelle data for the strongest emission lines (see Paper I). For
those lines with  S/N too low to perform a self-consistent
fitting of the different kinematical components present in their emission
profiles, the iterative process used to deconvolve the different components
did not yield meaningful results. For these weak emission lines we
used the fits found for the strongest emission lines {in Paper I} with
similar 
ionization degree as the initial guess as input for the task
\texttt{ngaussfit} of {\sc IRAF}. With these we fit the emission profiles of
the weak lines fixing the centroids and widths of the corresponding initial
approximation, allowing only the profiles amplitudes to vary.

We assumed a two ionization zone scheme to select the appropriate initial
approximation to fit the weak lines: the low ionization zone where [\OI],
[\OII], [\NII], [\SII], [\SIII] and [\ArIII] forbidden lines are originated,
and the high ionization zone 
where the helium recombination lines and the [\OIII] and [\NeIII] forbidden
lines are emitted. We used the {self-consistent} solution found in Paper I
for H$\alpha$, 
[\NII] or [\SII] (depending on the minimization of the fitting errors) in the
low ionization zone, and [\OIII] in the high ionization one.

\begin{figure}
\centering
\includegraphics[angle=0,width=.47\textwidth]{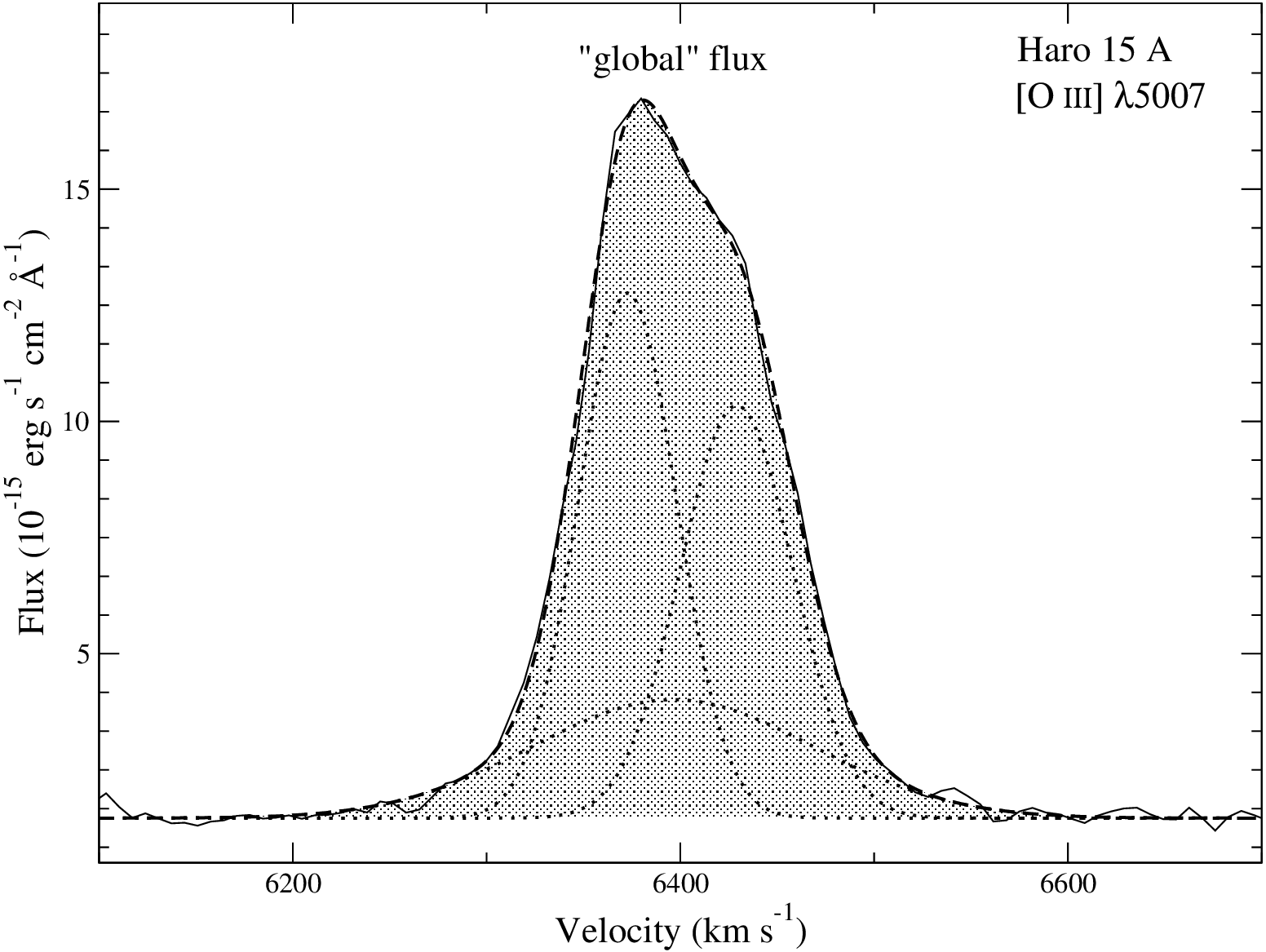}\\
\vspace{0.2cm}
\includegraphics[angle=0,width=.47\textwidth]{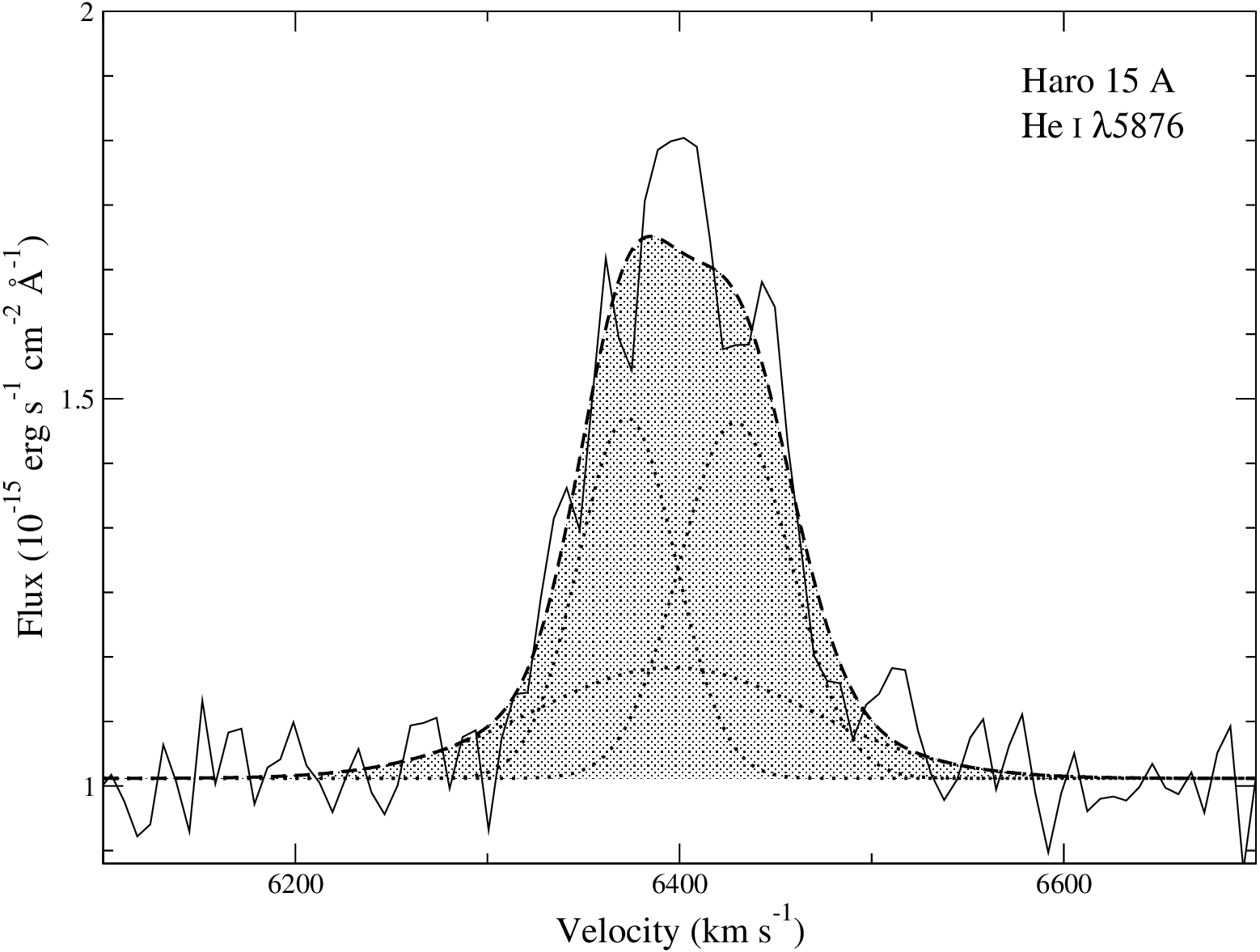}\\
\caption[]{[\OIII]\,$\lambda$\,5007\,\AA\ (upper panel) and
  \HeI\,$\lambda$\,5876\,\AA\ (lower panel) emission lines of knot A in the
  Flux-Velocity plane with their kinematical decomposition and the sum
  (hatched area) of the different components superposed.}
\label{fig:knotA}
\end{figure}

\begin{figure}
\centering
\includegraphics[angle=0,width=.47\textwidth]{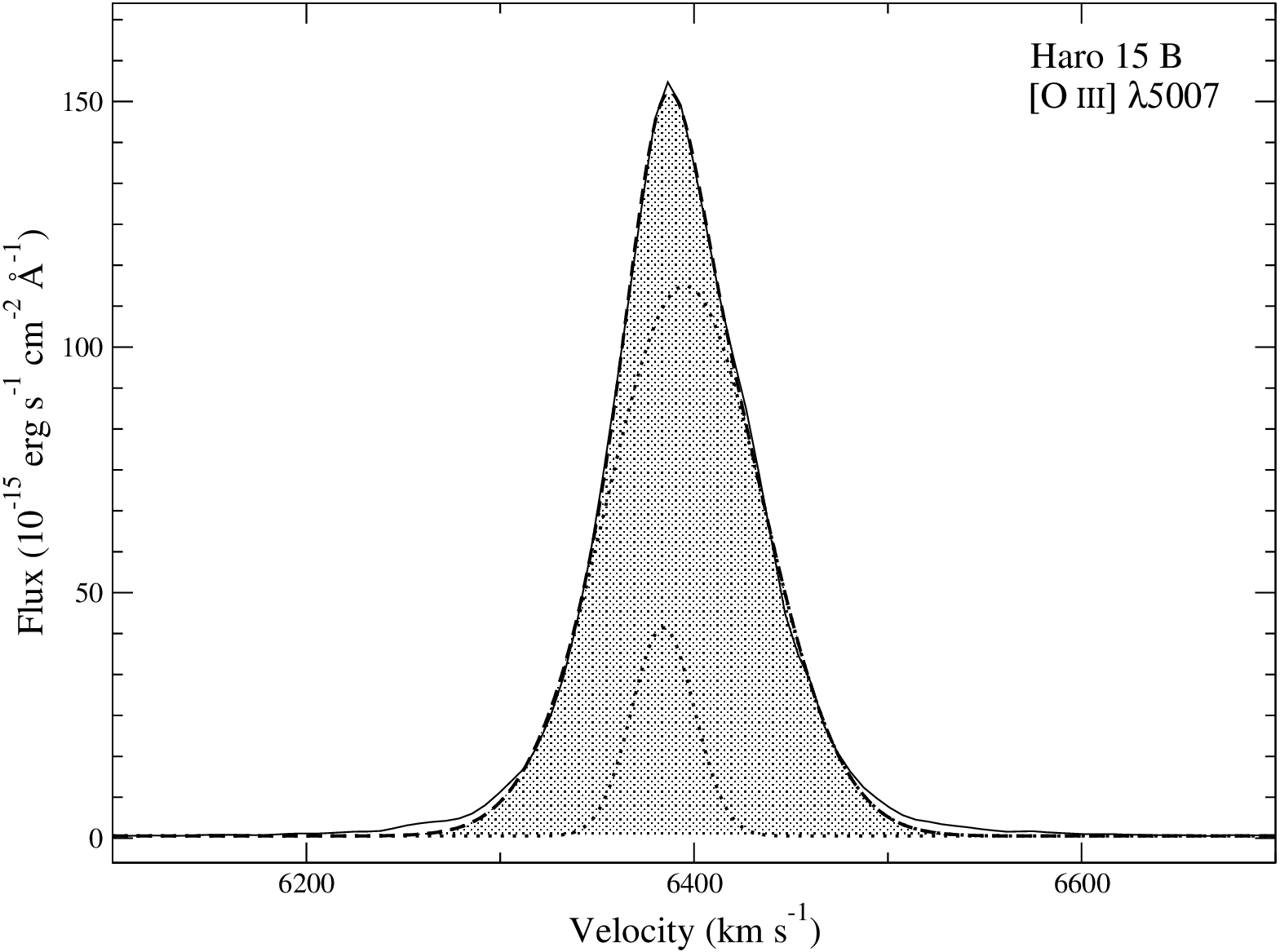}\\
\vspace{0.2cm}
\includegraphics[angle=0,width=.47\textwidth]{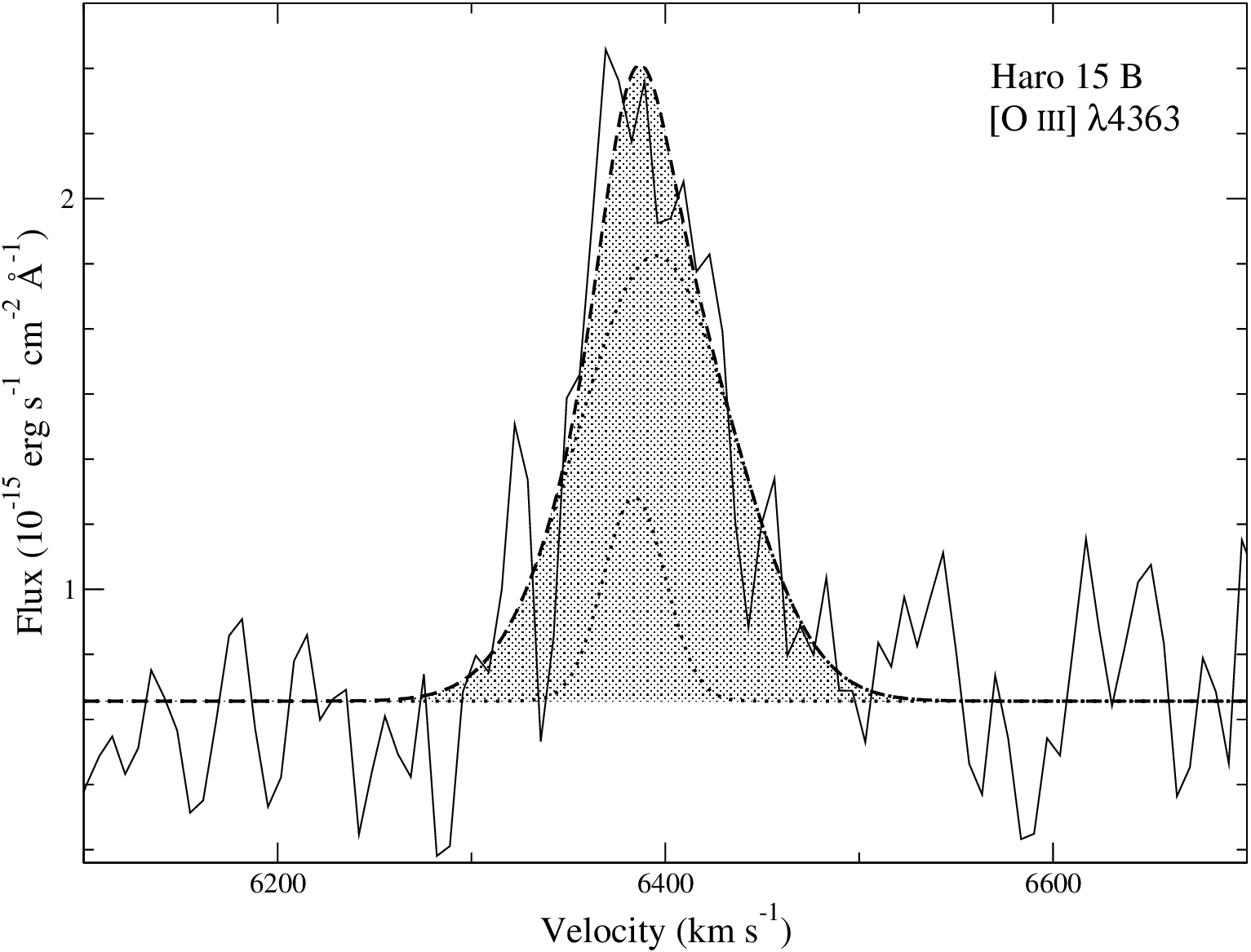}\\
\caption[]{[\OIII]\,$\lambda$\,5007\,\AA\ (upper panel) and
  [\OIII]\,$\lambda$\,4363\,\AA\ (lower panel) emission lines of knot B in the
  Flux-Velocity plane with their kinematical decomposition and the sum
  (hatched area) of the different components superposed.}
\label{fig:knotB}
\end{figure}

Upper panels of Figs.\ \ref{fig:knotA} and \ref{fig:knotB}  show the
[\OIII]\,$\lambda$\,5007\,\AA\ emission lines in the Flux-Velocity plane with
their kinematical components overlapped for knots A and B, respectively. 
In the lower panels of these figures, we show examples of the fittings performed
 for \HeI\,$\lambda$\,5876\,\AA\ and [\OIII]\,$\lambda$\,4363\,\AA\ for
knots A and B, respectively, using as the initial approximations those shown
in the upper panels. We can appreciate that this procedure for the weak
emission lines gives very good fittings with small residuals. 

The existence of an underlying stellar population is suggested by the detection
of absorption features that depress the Balmer emission lines in
the long-slit and echelle spectra. Therefore, the errors introduced by this
effect in the measurement of line intensities were minimized by
defining a pseudo-continuum  at the base of the hydrogen emission lines
(see H06). The presence of the wings of the
absorption lines imply that, even though we have used a pseudo-continuum,
there is still a fraction of the emitted flux that we are not able
to measure accurately \cite[see discussion in][]{1988MNRAS.231...57D}. This
fraction is not the same for all lines, nor are the ratios between the
absorbed fractions and the emission. In H06 it was
estimated that the difference in Balmer lines fluxes calculated using the
defined pseudo-continuum or a multi-Gaussian fit to the absorption and
emission components lies within the errors. At any
rate, for the Balmer emission lines we have doubled the estimated error,
$\sigma_l$, as a conservative approach to consider the uncertainties
introduced by the presence of the underlying stellar population.

The reddening coefficient [c(H$\beta$)] was calculated 
assuming the galactic extinction law of \cite{1972ApJ...172..593M} with
$R_{\rm v}$=3.2. We obtained a value for c(H$\beta$)  in each case by performing a
least-square fit to the observed ratio between F($\lambda_B$) and F(H$\beta$), where B denotes the different Balmer lines, to their theoretical values computed by \cite{1995MNRAS.272...41S} using an iterative
method to estimate $n_e$ and $T_e$ in each case \citep{2008PhDT........35H}.
Whenever available, initial $n_e$ and $T_e$ guesses were taken from measured
[\SII]~$\lambda\lambda$\,6717,6731 and [\OIII]~$\lambda\lambda$\,4363,4959,5007
line fluxes. We have considered $n_e$ equal to n([\SII]). Owing to the large
error introduced by the presence of the underlying stellar population, only
the four strongest Balmer emission lines (H$\alpha$, H$\beta$, H$\gamma$ and
H$\delta$) have been taken into account.

The emission line fluxes of the observed knots of \Haro\ are listed in Table
\ref{ratiostot 1} for the long-slit data, and in Tables \ref{ratiostot 2}-\ref{ratiostot 5} for the echelle data. Each table lists the reddening 
corrected emission lines ratios for each measurement, either for
global or kinematical component, together with the reddening constant and its
error taken as the uncertainties of the least square fit, and the reddening
corrected H$\beta$ intensity. First two columns share the same information in Tables \ref{ratiostot 1}-\ref{ratiostot 5}. Column 1 lists the wavelength and identification of
the measured lines. The adopted reddening curve, f($\lambda$), normalized to
H$\beta$, is given in column 2. The following columns show the equivalent widths (EW) in
\AA, the reddening corrected emission line intensities relative to H$\beta$, and their corresponding errors obtained by propagating in quadrature the observational errors in
the emission line fluxes and the reddening constant uncertainties for different knot/component accordingly. We have not taken into account errors in the theoretical intensities since they are lower than the observational ones.

\begin{table*}
\caption[]{Relative reddening corrected line intensities [$F(H\beta)$=$I(H\beta)$=10000] for the global measure of the long-slit spectrum of knots B and C.}
\label{ratiostot 1}
\begin{center}
\begin{tabular}{@{}l@{\hspace{0.1cm}}c@{\hspace{0.25cm}}c@{\hspace{0.25cm}}c@{\hspace{0.25cm}}c@{\hspace{0.25cm}}c@{\hspace{0.25cm}}c@{\hspace{0.25cm}}c@{}}
\hline
 &  & \multicolumn{3}{c}{B} & \multicolumn{3}{c}{C} \\
 \multicolumn{1}{c}{$\lambda$ ({\AA})} & f($\lambda$) & -EW  & $I(\lambda)$ & Error & -EW  & $I(\lambda)$ & Error  \\
  & & (\AA) & & (\%)  & (\AA) & & (\%)\\
  \hline
 3727 [\OII]$^a$       &   0.271   &  109.1 & 11940 $\pm$  120 &  1.0 &   45.3 & 25400 $\pm$ 2100 &  8.3 \\
 3835 H9              &   0.246   &    3.8 &   470 $\pm$   20 &  5.0 &  \ldots &  \ldots & \ldots \\
 3868 [\NeIII]         &   0.238   &   34.0 &  4490 $\pm$   10 &  0.2 &  \ldots &  \ldots & \ldots \\
 3889 \HeI+H8          &   0.233   &    9.4 &  1260 $\pm$   70 &  5.1 &  \ldots &  \ldots & \ldots \\
 3970 [\NeIII]+H$\epsilon$      &   0.215   &   21.1 &  2535 $\pm$    4 &  0.2 &  \ldots &  \ldots & \ldots \\
 4068 [\SII]           &   0.195   &    0.7 &    90 $\pm$   20 & 19.6 &  \ldots &  \ldots & \ldots \\
 4102 H$\delta$          &   0.188   &   21.1 &  2400 $\pm$   30 &  1.3 &  \ldots &  \ldots & \ldots \\
 4340 H$\gamma$          &   0.142   &   41.5 &  4480 $\pm$   20 &  0.4 &    6.0 &  3850 $\pm$  660 & 17.1 \\
 4363 [\OIII]          &   0.138   &    6.3 &   690 $\pm$   40 &  6.0 &  \ldots &  \ldots & \ldots \\
 4471 \HeI             &   0.106   &    2.8 &   310 $\pm$   20 &  5.2 &    3.8 &  2840 $\pm$  570 & 19.9 \\
 4658 [\FeIII]         &   0.053   &    0.3 &    36 $\pm$    6 & 17.8 &  \ldots &  \ldots & \ldots \\
 4686 \HeII            &   0.045   &    1.5 &   150 $\pm$   20 & 12.1 &    1.6 &  1260 $\pm$  330 & 26.1 \\
 4713 [\ArIV]+\HeI      &   0.038   &    0.6 &   64 $\pm$    7 & 11.4 &  \ldots &  \ldots & \ldots \\
 4740 [\ArIV]          &   0.031   &    0.9 &    90 $\pm$   10 & 14.1 &  \ldots &  \ldots & \ldots \\
 4861 H$\beta$           &   0.000   &  114.1 & 10000 $\pm$   30 &  0.3 &   14.6 & 10000 $\pm$  540 &  5.4 \\
 4959 [\OIII]          &  -0.024   &  221.3 & 21220 $\pm$   80 &  0.4 &    8.3 &  6680 $\pm$  210 &  3.1 \\
 5007 [\OIII]          &  -0.035   &  539.1 & 51340 $\pm$   90 &  0.2 &   27.9 & 22040 $\pm$  470 &  2.1 \\
 5876 \HeI             &  -0.209   &   22.5 &  1240 $\pm$   30 &  2.3 &  \ldots &  \ldots & \ldots \\
 6300 [\OI]            &  -0.276   &    6.8 &   370 $\pm$   20 &  4.3 &    2.5 &  1540 $\pm$  300 & 19.1 \\
 6312 [\SIII]          &  -0.278   &    2.3 &   120 $\pm$   20 & 15.5 &  \ldots &  \ldots & \ldots \\
 6364 [\OI]            &  -0.285   &    2.2 &   120 $\pm$   10 &  9.6 &  \ldots &  \ldots & \ldots \\
 6548 [\NII]           &  -0.311   &    8.3 &   420 $\pm$   20 &  4.8 &  \ldots &  \ldots & \ldots \\
 6563 H$\alpha$           &  -0.313   &  565.8 & 27620 $\pm$   40 &  0.1 &   52.6 & 28500 $\pm$  820 &  2.9 \\
 6584 [\NII]           &  -0.316   &   23.4 &  1180 $\pm$   20 &  1.8 &    5.8 &  3230 $\pm$  370 & 11.5 \\
 6678 \HeI             &  -0.329   &    5.5 &   260 $\pm$   10 &  4.7 &  \ldots &  \ldots & \ldots \\
 6717 [\SII]           &  -0.334   &   32.1 &  1450 $\pm$   20 &  1.7 &   10.2 &  5210 $\pm$  610 & 11.8 \\
 6731 [\SII]           &  -0.336   &   13.7 &   620 $\pm$   50 &  8.0 &    5.1 &  2610 $\pm$  490 & 18.9 \\
 7065 \HeI             &  -0.377   &    6.3 &   250 $\pm$   20 &  7.2 &  \ldots &  \ldots & \ldots \\
 7136 [\ArIII]         &  -0.385   &   19.4 &   710 $\pm$   20 &  3.0 &  \ldots &  \ldots & \ldots \\
 7281 \HeI$^b$         &  -0.402   &    1.1 &    42 $\pm$    5 & 11.9 &  \ldots &  \ldots & \ldots \\
 7319 [\OII]$^c$       &  -0.406   &    4.0 &   148 $\pm$    6 &  3.9 &  \ldots &  \ldots & \ldots \\
 7330 [\OII]$^d$       &  -0.407   &    3.1 &   116 $\pm$    4 &  3.2 &  \ldots &  \ldots & \ldots \\
 7751 [\ArIII]         &  -0.451   &    5.8 &   190 $\pm$   10 &  7.0 &  \ldots &  \ldots & \ldots \\
 9069 [\SIII]          &  -0.561   &   69.9 &  1200 $\pm$  120 &  9.8 &  \ldots &  \ldots & \ldots \\
  \hline
\multicolumn{2}{@{}l}{I(H$\beta$)(erg\,s$^{-1}$\,cm$^{-2}$)}   & \multicolumn{3}{c}{4.49\,$\times$\,10$^{-14}$}  & \multicolumn{3}{c}{0.11\,$\times$\,10$^{-14}$}  \\
c(H$\beta$) &     & \multicolumn{3}{c}{ 0.29 $\pm$ 0.01}& \multicolumn{3}{c}{ 0.24 $\pm$ 0.12 }\\
\hline
\multicolumn{8}{@{}l} {$^a$\,[\OII]~$\lambda\lambda$\,3726\,+\,3729; $^b$\,possibly blend with an unknown line}\\
\multicolumn{8}{@{}l} {$^c$\,[\OII]~$\lambda\lambda$\,7318\,+\,7320;  $^d$\,[\OII]~$\lambda\lambda$\,7330\,+\,7331.}
\end{tabular}
\end{center}
\end{table*}

\begin{table*}
\footnotesize
\caption{Relative reddening corrected line intensities
  [$F(H\beta)$=$I(H\beta)$=10000] for the global measure and different
  kinematical components of the echelle spectrum of knot A.}
\label{ratiostot 2}

\begin{center}
\begin{tabular}{@{}l@{\hspace{0.1cm}}c@{\hspace{0.25cm}}c@{\hspace{0.25cm}}c@{\hspace{0.25cm}}c@{\hspace{0.25cm}}c@{\hspace{0.25cm}}c@{\hspace{0.25cm}}c@{\hspace{0.25cm}}c@{\hspace{0.25cm}}c@{\hspace{0.25cm}}c@{\hspace{0.25cm}}c@{\hspace{0.25cm}}c@{\hspace{0.25cm}}c@{}}
\hline
& & \multicolumn{3}{c}{global} & \multicolumn{3}{c}{narrow\,1} & \multicolumn{3}{c}{narrow\,2} & \multicolumn{3}{c}{broad} \\
 \multicolumn{1}{c}{$\lambda$ ({\AA})} & f($\lambda$) & -EW  & $I(\lambda)$ & Error & -EW  & $I(\lambda)$ & Error  & -EW  & $I(\lambda)$ & Error & -EW  & $I(\lambda)$ & Error  \\
  & & (\AA) & & (\%) & (\AA) & & (\%) & (\AA) & & (\%) & (\AA) & & (\%)  \\
 \hline
   3727 [\OII]$^a$       &   0.271   &   12.8 & 30070 $\pm$ 1700 &  5.7 &    6.7 & 34390 $\pm$ 1570 &  4.6 &    2.5 & 37850 $\pm$ 4180 & 11.0 &    1.9 & 33500 $\pm$ 5980 & 17.9 \\
  3868 [\NeIII]         &   0.238   &    1.3 &  2900 $\pm$  290 & 10.1 &  \ldots &  \ldots & \ldots &  \ldots &  \ldots & \ldots &  \ldots &  \ldots & \ldots \\
 3888 H8+\HeI          &   0.233   &    0.8 &  1280 $\pm$  200 & 15.8 &  \ldots &  \ldots & \ldots &  \ldots &  \ldots & \ldots &  \ldots &  \ldots & \ldots \\
  3970 [\NeIII]+H$\epsilon$     &   0.215   &    1.7 &  1940 $\pm$  360 & 18.7 &  \ldots &  \ldots & \ldots &  \ldots &  \ldots & \ldots &  \ldots &  \ldots & \ldots \\
  4102 H$\delta$          &   0.188   &    5.4 &  5340 $\pm$ 1300 & 24.4 &  \ldots &  \ldots & \ldots &  \ldots &  \ldots & \ldots &  \ldots &  \ldots & \ldots \\
 4340 H$\gamma$          &   0.142   &    6.2 &  5140 $\pm$  280 &  5.4 &    2.2 &  4590 $\pm$  190 &  4.1 &    0.8 &  4210 $\pm$  550 & 13.0 &    0.4 &  2800 $\pm$  940 & 33.6 \\
  4471 \HeI             &   0.106   &    0.4 &   470 $\pm$   90 & 19.6 &  \ldots &  \ldots & \ldots &  \ldots &  \ldots & \ldots &  \ldots &  \ldots & \ldots \\
  4861 H$\beta$           &   0.000   &   12.9 & 10000 $\pm$  340 &  3.4 &    6.2 & 10000 $\pm$  160 &  1.6 &    2.9 & 10000 $\pm$  290 &  2.9 &    2.1 & 10000 $\pm$  590 &  5.9 \\
 4959 [\OIII]          &  -0.024   &    6.3 &  6300 $\pm$  120 &  2.0 &    2.3 &  4500 $\pm$  170 &  3.8 &    2.3 &  9240 $\pm$  440 &  4.8 &    1.4 &  7800 $\pm$  940 & 12.0 \\
 5007 [\OIII]          &  -0.035   &   22.4 & 19400 $\pm$  220 &  1.1 &    8.3 & 14190 $\pm$  160 &  1.1 &    7.5 & 26530 $\pm$  320 &  1.2 &    5.4 & 26720 $\pm$  980 &  3.7 \\
  5876 \HeI             &  -0.209   &    1.7 &   990 $\pm$   60 &  6.1 &    0.6 &   690 $\pm$   80 & 10.9 &    0.7 &  1350 $\pm$  140 & 10.5 &    0.5 &  1540 $\pm$  350 & 22.5 \\
 6300 [\OI]            &  -0.276   &    1.7 &   760 $\pm$   40 &  5.7 &  \ldots &  \ldots & \ldots &  \ldots &  \ldots & \ldots &  \ldots &  \ldots & \ldots \\
  6548 [\NII]           &  -0.311   &    4.2 &  1860 $\pm$   70 &  3.6 &    2.1 &  1910 $\pm$   80 &  4.3 &    1.1 &  1440 $\pm$  110 &  7.3 &    0.7 &  1650 $\pm$  450 & 27.3 \\
 6563 H$\alpha$           &  -0.313   &   74.0 & 30100 $\pm$  220 &  0.7 &   31.8 & 28500 $\pm$  140 &  0.5 &   22.1 & 28500 $\pm$  210 &  0.7 &   12.9 & 28500 $\pm$  660 &  2.3 \\
 6584 [\NII]           &  -0.316   &   13.3 &  6010 $\pm$  170 &  2.9 &    5.9 &  5570 $\pm$  130 &  2.4 &    3.5 &  4810 $\pm$  200 &  4.2 &    3.2 &  7430 $\pm$  690 &  9.3 \\
 6678 \HeI             &  -0.329   &    1.0 &   420 $\pm$   50 & 11.4 &  \ldots &  \ldots & \ldots &  \ldots &  \ldots & \ldots &  \ldots &  \ldots & \ldots \\
 6717 [\SII]           &  -0.334   &    8.2 &  3800 $\pm$  120 &  3.2 &    3.1 &  2940 $\pm$  100 &  3.5 &    1.4 &  1780 $\pm$  130 &  7.3 &    3.8 &  8530 $\pm$  840 &  9.9 \\
 6731 [\SII]           &  -0.336   &    8.3 &  3250 $\pm$  210 &  6.4 &    1.9 &  1890 $\pm$  120 &  6.5 &    1.1 &  1430 $\pm$  130 &  9.2 &    2.2 &  5250 $\pm$  670 & 12.7 \\
 7065 \HeI             &  -0.377   &    0.6 &   270 $\pm$   90 & 32.9 &  \ldots &  \ldots & \ldots &  \ldots &  \ldots & \ldots &  \ldots &  \ldots & \ldots \\
 7136 [\ArIII]         &  -0.385   &    2.1 &   860 $\pm$   50 &  6.1 &    0.7 &   580 $\pm$   80 & 14.5 &    0.5 &   520 $\pm$   80 & 16.1 &    0.7 &  1440 $\pm$  380 & 26.4 \\
  9069 [\SIII]          &  -0.561   &    4.4 &  2170 $\pm$  140 &  6.3 &    1.6 &  1800 $\pm$  210 & 11.5 &    1.5 &  1850 $\pm$  250 & 13.6 &    0.5 &  1260 $\pm$  390 & 31.0 \\
 9532 [\SIII]          &  -0.592   &    5.6 &  3110 $\pm$  250 &  8.1 &    1.1 &  1450 $\pm$  250 & 17.0 &    2.2 &  2860 $\pm$  330 & 11.6 &    1.7 &  4630 $\pm$ 1180 & 25.4 \\
\hline
\multicolumn{2}{@{}l}{I(H$\beta$)(erg\,s$^{-1}$\,cm$^{-2}$)}   & \multicolumn{3}{c}{1.56\,$\times$\,10$^{-14}$}  & \multicolumn{3}{c}{0.85\,$\times$\,10$^{-14}$}  & \multicolumn{3}{c}{0.39\,$\times$\,10$^{-14}$}  & \multicolumn{3}{c}{0.28\,$\times$\,10$^{-14}$}  \\
c(H$\beta$) &     & \multicolumn{3}{c}{ 0.22 $\pm$ 0.04 }    & \multicolumn{3}{c}{ 0.11 $\pm$ 0.03 }    & \multicolumn{3}{c}{ 0.68 $\pm$ 0.05 }    & \multicolumn{3}{c}{ 0.37 $\pm$ 0.12 }  \\
\hline
\multicolumn{14}{@{}l}{$^a$\,[\OII]~$\lambda\lambda$\,3726\,+\,3729}
\end{tabular}
\end{center}
\end{table*}

\begin{table*} 
\caption[]{Relative reddening corrected line intensities [$F(H\beta)$=$I(H\beta)$=10000] for the global measure and different kinematical components of the echelle spectrum of knot B.}
\label{ratiostot 3}
{\footnotesize
\begin{center}
\begin{tabular}{@{}l@{\hspace{0.1cm}}c@{\hspace{0.25cm}}c@{\hspace{0.25cm}}c@{\hspace{0.25cm}}c@{\hspace{0.25cm}}c@{\hspace{0.25cm}}c@{\hspace{0.25cm}}c@{\hspace{0.25cm}}c@{\hspace{0.25cm}}c@{\hspace{0.25cm}}c@{}}
\hline
&  & \multicolumn{3}{c}{global} & \multicolumn{3}{c}{narrow} & \multicolumn{3}{c}{broad} \\
 \multicolumn{1}{c}{$\lambda$ ({\AA})} & f($\lambda$)  & -EW  & $I(\lambda)$ & Error & -EW  & $I(\lambda)$ & Error  & -EW  & $I(\lambda)$ & Error  \\
  & & (\AA) & & (\%) & (\AA) & & (\%) & (\AA) & & (\%)  \\
  \hline
 3727 [\OII]$^a$       &   0.271   &   20.4 &  9050 $\pm$  240 &  2.6 &    8.1 & 24100 $\pm$  850 &  3.5 &    8.0 &  5880 $\pm$  310 &  5.3 \\
 3835 H9              &   0.246   &    1.0 &   360 $\pm$   20 &  4.4 &  \ldots &  \ldots & \ldots &  \ldots &  \ldots & \ldots \\
 3868 [\NeIII]         &   0.238   &   12.4 &  5590 $\pm$  180 &  3.3 &    1.6 &  4210 $\pm$  650 & 15.4 &    8.4 &  5850 $\pm$  270 &  4.6 \\
 3889 \HeI+H8          &   0.233   &    5.1 &  1570 $\pm$   40 &  2.6 &  \ldots &  \ldots & \ldots &  \ldots &  \ldots & \ldots \\
 3968 [\NeIII]+H7      &   0.216   &    5.2 &  1500 $\pm$   30 &  2.0 &    0.5 &  1020 $\pm$  380 & 36.9 &    2.8 &  1500 $\pm$  160 & 10.8 \\
 3970 [\NeIII]+H$\epsilon$     &   0.215   &    3.8 &  1060 $\pm$   60 &  5.3 &    1.2 &  1630 $\pm$  240 & 14.9 &    3.4 &  1260 $\pm$  100 &  8.2 \\
 4102 H$\delta$          &   0.188   &    9.7 &  2770 $\pm$  120 &  4.3 &    1.4 &  2220 $\pm$  340 & 15.2 &    5.7 &  2530 $\pm$  110 &  4.5 \\
 4340 H$\gamma$          &   0.142   &   37.8 &  4830 $\pm$  120 &  2.5 &    5.6 &  5310 $\pm$  160 &  3.0 &   16.7 &  4430 $\pm$   60 &  1.3 \\
 4363 [\OIII]          &   0.138   &    4.4 &   780 $\pm$  130 & 16.3 &    0.4 &   470 $\pm$  210 & 45.8 &    2.4 &   720 $\pm$   90 & 12.5 \\
 4471 \HeI             &   0.106   &    3.0 &   470 $\pm$   20 &  3.6 &  \ldots &  \ldots & \ldots &  \ldots &  \ldots & \ldots \\
 4713 [\ArIV]+\HeI      &   0.038   &    2.6 &   340 $\pm$   30 &  8.4 &  \ldots &  \ldots & \ldots &  \ldots &  \ldots & \ldots \\
 4861 H$\beta$           &   0.000   &   81.3 & 10000 $\pm$ 100 &  1.0 &   14.1 & 10000 $\pm$  160 &  1.6 &   45.6 & 10000 $\pm$   70 &  0.7 \\
 4959 [\OIII]          &  -0.024   &  130.0 & 22710 $\pm$  180 &  0.8 &   16.7 & 14960 $\pm$  290 &  1.9 &   91.0 & 25570 $\pm$  150 &  0.6 \\
 5007 [\OIII]          &  -0.035   &  470.0 & 69810 $\pm$  210 &  0.3 &   35.0 & 37510 $\pm$  800 &  2.1 &  233.8 & 79550 $\pm$  340 &  0.4 \\
 5876 \HeI             &  -0.209   &   12.4 &  1053 $\pm$    9 &  0.9 &    1.7 &   570 $\pm$   40 &  6.5 &   10.2 &  1240 $\pm$   20 &  1.6 \\
 6300 [OI]            &  -0.276   &    2.7 &   181 $\pm$    8 &  4.4 &    2.0 &   500 $\pm$   40 &  7.5 &    0.8 &    70 $\pm$   20 & 30.3 \\
 6312 [\SIII]          &  -0.278   &    1.8 &   130 $\pm$   30 & 20.2 &    0.7 &   200 $\pm$   40 & 22.2 &    1.0 &   100 $\pm$   30 & 24.1 \\
 6548 [\NII]           &  -0.311   &    2.4 &   185 $\pm$    5 &  2.6 &    1.6 &   420 $\pm$   20 &  5.7 &    1.6 &   160 $\pm$   20 & 13.0 \\
 6563 H$\alpha$           &  -0.313   &  358.0 & 28620 $\pm$   40 &  0.1 &   73.5 & 29350 $\pm$  270 &  0.9 &  184.8 & 28010 $\pm$  130 &  0.5 \\
 6584 [\NII]           &  -0.316   &    7.7 &   620 $\pm$   10 &  1.9 &    6.0 &  1570 $\pm$   50 &  3.4 &    4.0 &   400 $\pm$   30 &  6.2 \\
 6678 \HeI             &  -0.329   &    4.8 &   320 $\pm$    3 &  1.1 &    0.5 &   130 $\pm$   30 & 20.7 &    4.6 &   430 $\pm$   20 &  3.8 \\
 6717 [\SII]           &  -0.334   &   10.6 &   730 $\pm$   10 &  1.3 &    7.4 &  1920 $\pm$   50 &  2.4 &    3.6 &   360 $\pm$   30 &  8.1 \\
 6731 [\SII]           &  -0.336   &    8.4 &   590 $\pm$   20 &  3.8 &    5.2 &  1410 $\pm$   50 &  3.8 &    2.6 &   270 $\pm$   30 & 12.6 \\
 7065 \HeI             &  -0.377   &    4.9 &   290 $\pm$    6 &  1.9 &  \ldots &  \ldots & \ldots &  \ldots &  \ldots & \ldots \\
 7136 [\ArIII]         &  -0.385   &   11.0 &   710 $\pm$    9 &  1.3 &    4.0 &   980 $\pm$   50 &  5.3 &    6.9 &   670 $\pm$   20 &  3.4 \\
 7751 [\ArIII]         &  -0.451   &    3.2 &   200 $\pm$   10 &  6.4 &  \ldots &  \ldots & \ldots &  \ldots &  \ldots & \ldots \\
 9069 [\SIII]          &  -0.561   &   13.9 &  1190 $\pm$  110 &  9.3 &    3.1 &  1010 $\pm$  140 & 13.5 &    9.2 &  1330 $\pm$   90 &  6.8 \\
 9532 [\SIII]          &  -0.592   &   21.3 &  2510 $\pm$  140 &  5.6 &    6.9 &  2890 $\pm$  190 &  6.4 &   13.1 &  2490 $\pm$  130 &  5.0 \\
\hline
\multicolumn{2}{@{}l}{I(H$\beta$)(erg\,s$^{-1}$\,cm$^{-2}$)}   & \multicolumn{3}{c}{3.16\,$\times$\,10$^{-14}$}  & \multicolumn{3}{c}{0.74\,$\times$\,10$^{-14}$}  & \multicolumn{3}{c}{2.40\,$\times$\,10$^{-14}$}  \\
c(H$\beta$) &     & \multicolumn{3}{c}{ 0.13 $\pm$ 0.01 }    & \multicolumn{3}{c}{ 0.36 $\pm$ 0.02 }    & \multicolumn{3}{c}{ 0.07 $\pm$ 0.01 }  \\
\hline
\multicolumn{11}{@{}l}{$^a$\,[\OII]~$\lambda\lambda$\,3726\,+\,3729}
\end{tabular}
\end{center}}
\end{table*}

\begin{table*} 
{\footnotesize
\caption[]{Relative reddening corrected line intensities [$F(H\beta)$=$I(H\beta)$=10000] for the global measure and different kinematical components of the echelle spectrum of knot C.}
\label{ratiostot 4}
\begin{center}
\begin{tabular}{@{}l@{\hspace{0.1cm}}c@{\hspace{0.25cm}}c@{\hspace{0.25cm}}c@{\hspace{0.25cm}}c@{\hspace{0.25cm}}c@{\hspace{0.25cm}}c@{\hspace{0.25cm}}c@{\hspace{0.25cm}}c@{\hspace{0.25cm}}c@{\hspace{0.25cm}}c@{}}
\hline
&  & \multicolumn{3}{c}{global} & \multicolumn{3}{c}{narrow} & \multicolumn{3}{c}{broad} \\
 \multicolumn{1}{c}{$\lambda$ ({\AA})} & f($\lambda$) & -EW  & $I(\lambda)$ & Error & -EW  & $I(\lambda)$ & Error  & -EW  & $I(\lambda)$ & Error  \\
  & & (\AA) & & (\%) & (\AA) & & (\%) & (\AA) & & (\%)  \\
  \hline
 4861 H$\beta$           &   0.000   &    3.2 & 10000 $\pm$ 2190 & 21.9 &    1.0 & 10000 $\pm$ 1460 & 14.6 &    1.1 & 10000 $\pm$ 2350 & 23.5 \\
 4959 [\OIII]          &  -0.024   &    5.6 &  6880 $\pm$  390 &  5.7 &    3.4 &  9820 $\pm$  510 &  5.2 &    2.0 &  5250 $\pm$  610 & 11.7 \\
 5007 [\OIII]          &  -0.035   &    5.6 & 20470 $\pm$ 1270 &  6.2 &    3.4 & 29370 $\pm$ 1580 &  5.4 &    2.0 & 15580 $\pm$ 1860 & 11.9 \\
 6563 H$\alpha$           &  -0.313   &   18.1 & 28500 $\pm$ 1150 &  4.0 &    6.5 & 28500 $\pm$  710 &  2.5 &    8.0 & 28500 $\pm$  800 &  2.8 \\
 6584 [\NII]           &  -0.316   &    1.3 &  2330 $\pm$  740 & 31.7 &    0.9 &  3850 $\pm$  940 & 24.4 &    0.4 &  1410 $\pm$  590 & 41.7 \\
  6717 [\SII]           &  -0.334   &    2.6 &  5020 $\pm$ 1590 & 31.7 &    1.2 &  5400 $\pm$ 1360 & 25.1 &    1.4 &  5320 $\pm$ 1890 & 35.6 \\
 6731 [\SII]           &  -0.336   &    1.8 &  2890 $\pm$ 1070 & 37.0 &    0.7 &  3190 $\pm$ 1300 & 40.6 &    1.0 &  3890 $\pm$ 1610 & 41.3 \\
  \hline
\multicolumn{2}{@{}l}{I(H$\beta$)(erg\,s$^{-1}$\,cm$^{-2}$)}   & \multicolumn{3}{c}{0.08\,$\times$\,10$^{-14}$}  & \multicolumn{3}{c}{0.04\,$\times$\,10$^{-14}$}  & \multicolumn{3}{c}{0.04\,$\times$\,10$^{-14}$}  \\
     c(H$\beta$) &     & \multicolumn{3}{c}{ 0.21 $\pm$ 0.40 }    & \multicolumn{3}{c}{ 0.14 $\pm$ 0.25 }    & \multicolumn{3}{c}{ 0.38 $\pm$ 0.41 }  \\
  \hline
\end{tabular}
\end{center}}
\end{table*}

\begin{table*} 
 {\footnotesize
\caption[]{Relative reddening corrected line intensities [$F(H\beta)$=$I(H\beta)$=10000] for the global measure and different kinematical components of the echelle spectrum of knot E.} 
\label{ratiostot 5}
\begin{center}
\begin{tabular}{@{}l@{\hspace{0.1cm}}c@{\hspace{0.25cm}}c@{\hspace{0.25cm}}c@{\hspace{0.25cm}}c@{\hspace{0.25cm}}c@{\hspace{0.25cm}}c@{\hspace{0.25cm}}c@{\hspace{0.25cm}}c@{\hspace{0.25cm}}c@{\hspace{0.25cm}}c@{}}
\hline
 &  & \multicolumn{3}{c}{global} & \multicolumn{3}{c}{narrow\,1} & \multicolumn{3}{c}{narrow\,2} \\
 \multicolumn{1}{c}{$\lambda$ ({\AA})} & f($\lambda$) & -EW  & $I(\lambda)$ & Error& -EW  & $I(\lambda)$ & Error  & -EW  & $I(\lambda)$ & Error\\
  & & (\AA) & & (\%)  & (\AA) & & (\%)& (\AA) & & (\%)\\
  \hline
 3727 [\OII]$^a$       &   0.271   &    8.2 & 32110 $\pm$ 9750 & 30.3 &  \ldots & \ldots &\ldots &  \ldots & \ldots &\ldots \\
 4861 H$\beta$           &   0.000   &    7.5 & 10000 $\pm$ 1530 & 15.3 &    1.8 & 10000 $\pm$ 1270 & 12.7 &    1.9 & 10000 $\pm$ 1800 & 18.2 \\
 4959 [\OIII]          &  -0.024   &    2.9 &  5750 $\pm$  730 & 12.7 &    2.2 &  3440 $\pm$  350 & 10.2 &    4.4 &  6900 $\pm$  380 &  5.6 \\
 5007 [\OIII]          &  -0.035   &    6.9 & 12160 $\pm$  920 &  7.6 &    2.2 & 10160 $\pm$ 1050 & 10.3 &    4.4 & 20670 $\pm$ 1210 &  5.9 \\
 5876 \HeI             &  -0.209   &    1.6 &  1730 $\pm$  560 & 32.4 &  \ldots & \ldots &\ldots &  \ldots & \ldots &\ldots \\
 6300 [OI]            &  -0.276   &    2.1 &  2020 $\pm$  560 & 27.9 &  \ldots & \ldots &\ldots &  \ldots & \ldots &\ldots \\
 6548 [\NII]           &  -0.311   &    1.4 &  1420 $\pm$  470 & 33.0 &  \ldots & \ldots &\ldots &  \ldots & \ldots &\ldots \\
 6563 H$\alpha$           &  -0.313   &   32.0 & 28500 $\pm$ 1090 &  3.8 &   15.9 & 28500 $\pm$  640 &  2.2 &   11.0 & 28500 $\pm$  960 &  3.4 \\
 6584 [\NII]           &  -0.316   &    5.8 &  5270 $\pm$ 1160 & 22.0 &    1.7 &  2810 $\pm$  620 & 22.0 &    3.7 &  9030 $\pm$ 2220 & 24.6 \\
 6717 [\SII]           &  -0.334   &    6.0 &  6210 $\pm$ 1440 & 23.2 &    1.4 &  2510 $\pm$  510 & 20.4 &    4.3 & 11230 $\pm$ 2880 & 25.6 \\
 6731 [\SII]           &  -0.336   &    5.2 &  4450 $\pm$ 1400 & 31.5 &    0.8 &  1880 $\pm$  790 & 42.1 &    1.8 &  6640 $\pm$ 2630 & 39.6 \\
  \hline
\multicolumn{2}{@{}l}{I(H$\beta$)(erg\,s$^{-1}$\,cm$^{-2}$)}   & \multicolumn{3}{c}{0.18\,$\times$\,10$^{-14}$}  & \multicolumn{3}{c}{0.07\,$\times$\,10$^{-14}$}  & \multicolumn{3}{c}{0.07\,$\times$\,10$^{-14}$}  \\
     c(H$\beta$) &     & \multicolumn{3}{c}{ 0.00 }    & \multicolumn{3}{c}{ 0.59 $\pm$ 0.22 }    & \multicolumn{3}{c}{0.00 }  \\
 \hline
     \multicolumn{11}{@{}l}{$^a$\,[\OII]~$\lambda\lambda$\,3726\,+\,3729}
\end{tabular}
\end{center}}
\end{table*}


As the  [\SIII]\,$\lambda$\,9532\,\AA\ line is outside the spectral range
of the long-slit data, all the physical parameters of knot B that depend on
this emission line were calculated using the theoretical ratio between this
line and [\SIII]\,$\lambda$\,9069\AA, I(9532)$\approx$2.44$\times$I(9069) 
\citep{O89}. In the case of knot C, we were not able to measure the emission
line flux of [\SIII]\,$\lambda$\,9069\AA\ with acceptable accuracy. Thus, we
do not include this line either in the table or in our calculations. For this
knot, it was impossible to measure the [\NII]\,$\lambda$\,6548\,\AA\ emission line
since it is blended with H$\alpha$. We corrected the H$\alpha$ flux for
the contamination by this nitrogen line estimating this contribution from
[\NII]\,$\lambda$\,6584\,\AA\ using the the theoretical ratio,
I(6584)\,$\approx$\,2.9$\,\cdot$\,I(6548) \citep{O89}. 

We can appreciate in the echelle data that the
[\SII]\,$\lambda$\,6731\,\AA\ emission line is affected 
by OH sky telluric emission lines present in the same spectral range due to
the galaxy redshift. We used the standard star Feige\,110 (sdOB) as telluric
star since this star has no stellar lines in this spectral range. Using {\sc
  IRAF} \texttt{telluric} routing, we have corrected the echelle 
spectra by this effect. The following step was to resort to the
\texttt{ngaussfit} task using the fitting solution for
[\SII]\,$\lambda$\,6717\,\AA\ as the initial approximation to fit the
[\SII]\,$\lambda$\,6731\,\AA\ line profile fixing the 
centroids and widths of the components, allowing only to fit the amplitudes.  

The emission lines with relative flux errors higher than 45\% were not taken
into account for the calculations. The exception is
[\OIII]\,$\lambda$\,4363\,\AA\ auroral emission line in knot B, which, albeit
faint, is essential to estimate the [\OIII] line temperature. 
 We were not able to measure the
[\OII]\,$\lambda\lambda$\,7319,7330\,\AA\ auroral emission lines for any
echelle spectra. In knots C 
and E, the [\SIII]$\lambda$\,9069,9532\,\AA\ lines were not detected due to the
low signal-to-noise of these spectra in this spectral range. In these knots
the [\OIII]\,$\lambda$\,4959\,\AA\ emission line fluxes were calculated from the
[\OIII]\,$\lambda$\,5007\,\AA\ line using the theoretical ratio,
I(5007)\,$\approx$\,3$\times$\,I(4959) \citep{O89}. In knot A, the weak
H$\delta$ emission line was not considered for calculations. For knot E, the
narrow\,2 component and the global measure present c(H$\beta$) values slightly
lower than zero but still compatible with this value within the observational
errors. In this case we considered c(H$\beta$) as zero. 


\subsection{Physical conditions of the ionized gas}
\label{sec:PhCond} 

We have studied the physical conditions of the emitting medium in four
star-forming knots of \Haro\ using long-slit (only for knots B and C) and
echelle data. With the echelle data, we have analysed the conditions in each knot
from the global and kinematical decomposition measurements following the
results derived in Paper I. In what follows we will explain the different
methods used to estimate the electron temperatures ($T_e$) and densities
[N$_e$\,$\approx$\,n([\SII])] \cite[see discussion in][and 
references therein]{FirpoPhDT}. 

\begin{enumerate}[I]
\item When the weak auroral lines needed to
implement the direct method could be measured we have derived the physical
conditions using the direct method as described in H08 and H\"agele et
al.\ (\citeyear[hereafter H11]{2011MNRAS.tmp..430H}). This procedure is 
based on the five-level statistical equilibrium atom approximation in the task
\texttt{temden}, of the software package {\sc IRAF}
\citep{1987JRASC..81..195D,1995PASP..107..896S}.  We have
taken the uncertainties associated with the measurement of the emission-line
fluxes and the reddening correction as error sources, and we have propagated them through our
calculations. The adopted collisional atomic coefficients are the same used in
H08. 

\item When some of the weak auroral lines are not detected in the spectrum of an
observed region, and the line temperatures cannot be computed using the direct
method, we have used relations between temperatures derived using
photoionization models \citep{2003MNRAS.346..105P,2005MNRAS.361.1063P}, or
empirical and semi-empirical relations
\citep{2006MNRAS.372..293H,2007MNRAS.382..251D,2007MNRAS.375..685P}. 
\cite{2003MNRAS.346..105P,2005MNRAS.361.1063P} examined some relationships
between temperatures used for ionic abundance determinations, adapting these
relations to the latest photoionization model results at that moment and using
the most recent atomic coefficients. We have to stress that the errors of the
line temperatures derived using these model dependent relations are formal
errors calculated from the measured line intensity errors applying error
propagation formulae, without assigning any error to the temperature
calibration itself. We have estimated line temperatures following the method described in H06 and H08 which takes advantage of the temperatures derived using the direct method in combination with either model based or empirical relations.

\item When no auroral emission line was detected in the spectrum, we have used
empirical methods to estimate the electron temperatures: 

\begin{itemize}
\item If the strong emission lines of [\SII], [\SIII], [\OII], and [\OIII] are
available in the spectrum of the studied region, we can use the relation
between the empirical parameter SO$_{23}$ \citep[=\,S$_{23}$/O$_{23}$; defined
  by][]{2000MNRAS.312..130D} and the [\SIII] line temperature
developed by D\'{\i}az et al.\ (\citeyear[hereafter D07]{2007MNRAS.382..251D}):
\[
t_e([S{\textsc{iii}}])\,=\,0.596\,-\,0.283\,\log\,SO_{23}\,+\,0.199\,(\log\,SO_{23})^{2}
\] 
\noindent 
This relationship was derived for high metallicity regions. However, from
Fig.\ 9 of D07, we can appreciate that this relation is also valid for the
H\,{\sc ii} galaxies in the upper end of the SO$_{23}$ distribution for this kind of
objects (where their dispersion is the lowest) and whose derived values of
SO$_{23}$ are in the validity range of the relation. Fortunately this is the
case for the SO$_{23}$ empirical parameter as measured in the star-forming knots of \Haro.

We used this T$_e$([\SIII]) to estimate T$_e$([\OIII]) using the
relation between these two line temperatures derived by H06: 
\[
t_e([S{\textsc{iii}}])\,=\,(1.19\,\pm\,0.08)\,t_e([O{\textsc{iii}}])\,-\,(0.32\,\pm\,0.10)
\] 
\noindent The temperature errors have been calculated from the
measured line intensity errors, 
taking also into account the temperature calibration errors.

\item If the [\SIII] emission lines are not detected with enough quality, we can
use the relation between the [\NII] nebular-to-auroral line intensities
as a function of the intensities of the strong nebular oxygen lines for the
determination of the [\NII] electron temperature (t$_2$) 
applying the empirical relation by Pilyugin (\citeyear[hereafter P07]{2007MNRAS.375..685P}):
\[
t_2\,=\,\frac{1.111}{\log Q_{N{\textsc{ii}}}\,-\,0.892\,-\,0.144\,\cdot\,\log\,t_2\,+\,0.023\,\cdot\,t_2}
\]
\noindent where Q$_{N{\textsc{ii}}}$ is the approximation to the [\NII] line
intensities ratio (=\,[\NII]\,$\lambda\lambda$\,6548,84/[\NII]\,$\lambda$\,5755). 
Q$_{N{\textsc{ii}}}$ depends on the strong
[\OII]\,$\lambda\lambda$\,3727,29\,\AA\ and
[\OIII]\,$\lambda\lambda$\,4959,5007\,\AA\ emission
lines. As the calibration of Pilyugin's method has been derived using data on high metallicity regions and its validity could be questioned in the low metallicity regime, we confirmed it in those regions were both methods could be applied, finding a good agreement.
It is therefore possible to estimate the line temperature of [\NII], which apparently
characterizes the low ionization area of the nebula, using the Pilyugin's
method. 
D07 found a systematic difference of about 500\,K between the [\NII] line temperature
using the Pilyugin's method and [\SIII] temperatures derived using their
empirical relation. This difference was accounted for when these determinations
were present in our work.
\end{itemize}

\item When we cannot measure either any auroral
emission line, or any intense lines to derive the
electron temperatures from empirical methods, for example in the two weakest star-forming knots, we considered T$_e$([\OIII])
equal to 10$^{4}$\,K, since it is a typical value for this kind of
objects. Theoretical and empirical relations were therefore used to derive the
other line temperatures from the adopted value for T$_e$([\OIII]).
\end{enumerate}

\subsection{Chemical abundance derivation}

To study the global and kinematical component abundances in each knot, we have
derived the ionic chemical abundances of different species. We have used the
strongest available emission lines detected in the analysed long-slit and
echelle spectra and the
task \texttt{ionic} of the STSDAS package in {\sc IRAF}, based on the
five-level statistical equilibrium atom approximation, as described in
H08. 

The total abundances have been derived taking into account, when required, the
unseen ionization stages of each element, resorting to the most widely
accepted ionization correction factors (ICFs) for each species,
X/H\,=\,ICF(X$^{+i}$)\,X$^{+i}$/H${^+}$
\cite[see][]{2007MNRAS.381..125P}. The procedure is
detailed in the following subsections.

\subsubsection{Helium}

The helium recombination lines arise mainly from pure recombination, although
they could have some contribution from collisional excitation and be affected
by self-absorption \citep[see][for a complete treatment of these
effects]{2001NewA....6..119O,2004ApJ...617...29O}. The electron temperature
T$_e$([\OIII]) is taken as representative of the zone where the He emission
arises since ratios of recombination lines are only weakly sensitive to
electron temperature. We have used the equations given by Olive and Skillman to
derive the He$^{+}$/H$^{+}$ value, using the theoretical emissivities scaled
to H$\beta$ from \cite{1999ApJ...514..307B} and the expressions for the
collisional correction factors from \cite{1995ApJ...442..714K}. To calculate
the abundance of twice ionized helium we have used equation (9) from
\cite{Kunth+83}. A summary of the equations used to calculate these ionic
abundances is given in Appendix B of \cite{2009PhDT........15G}.

It is possible to calculate the abundances of once and twice ionized helium using
 the \HeI~$\lambda\lambda$\,4471,5876,6678,7065\,\AA\ and
 \HeII~$\lambda$\,4686\,\AA\ 
emission lines, respectively. When these lines are available, we have
used them to derive the corresponding ionic abundance. As the observed
objects have low densities, three of the used helium lines have a small
dependence with optical depth effects, then we have not made any corrections
for the fluorescence. We have not corrected either for the presence of an
underlying stellar population. The total abundance of He has been found by
adding directly the two ionic abundances,
He/H\,=\,(He${^+}$+He$^{2+}$)/H$^+$. The results obtained for each line and 
the total He abundances, along with their corresponding errors are presented
in Table \ref{abundancesBC} and \ref{abundancesABCEF} for the long-slit and
echelle data, respectively. These tables also include the adopted value for
He$^{+}$/H$^{+}$ as the average value, weighted by the errors, of the abundances
derived from each \HeI\ emission line.  

\subsubsection{Ionic and total chemical abundances from forbidden lines}
\label{sec:abundances}

We have calculated the ionic and total abundances of O, S, N, Ne, and Ar
using the estimated line temperatures as described in H08. {The Fe ionic 
abundance has been derived using equations (11) and (12) from
\cite{Izotov+94}}. In those cases where there are no measurements 
of the [\SII]\,$\lambda$\,4068\,\AA\ and
[\OII]\,$\lambda$\,7325\,\AA\ auroral emission lines, and under the assumption
of an homogeneous electron temperature in the low-ionization zone, we have
taken the approximation T$_e$([\SII])$\approx$T$_e$([\OII])
\citep{2003MNRAS.346..105P}, where the [\OII] temperature were derived from
the [\OIII] ones using the relations given by photoionization models 
\citep{2003MNRAS.346..105P}. The nitrogen auroral line,
[\NII]\,$\lambda$\,5755\,\AA, is unobservable for all the studied star-forming
knots. Then, under the same assumption, we have estimated this temperature
using the approximation T$_e$([\NII])$\approx$T$_e$([\OII]), except in the
cases where we had to derive the [\NII] temperatures first using Pilyugin's
method and estimate the [\OII] ones using the same approximation.  

\begin{description}
\item[(i)] The oxygen ionic abundance ratios, O$^{+}$/H$^{+}$ and
O$^{2+}$/H$^{+}$, have been derived from the
[\OII]\,$\lambda\lambda$\,3727,29\,\AA\ and [\OIII]\,$\lambda\lambda$\,4959,
5007\,\AA\ lines, respectively, using for each ion its corresponding
temperature. At the temperatures derived for the observed objects, most of the
oxygen is in the form of O$^+$ and O$^{2+}$; therefore the approximation
O/H\,=\,(O$^+$\,+\,O$^{2+}$)/H$^+$ is valid to represent the total oxygen
abundances. 

\item[(ii)] The sulphur ionic abundances, S$^+$/H$^{+}$ and S$^{2+}$/H$^{+}$,
have been derived using T$_e$([\SII]) and T$_e$([\SIII]), and the fluxes of
the [\SII]\,$\lambda\lambda$\,6717,6731\,\AA\ and the near-IR
[\SIII]\,$\lambda\lambda$\,9069,9532\,\AA\ emission lines, respectively. A
relatively important contribution from S$^{3+}$ may be expected for sulphur,
depending on the nebular excitation. Taking into account this unseen
ionization state, the total sulphur abundance is calculated using an ICF for
S$^+$\,+\,S$^{2+}$. A good approximation for this ICF in terms of O$^+$/O is the
formula by \cite{1980ApJ...240...99B}, is based on \cite{1978A&A....66..257S}
photoionization models, with $\alpha$\,=\,2.5, which gives the best fit to the
scarce observational data on S$^{3+}$ abundances \citep{2006A&A...449..193P}. 
H08 further reduced the propagated error by writing this ICF in terms of the
ratio O$^{2+}$/O instead of O$^+$/O.

\item[(iii)]The nitrogen ionic abundance, N$^{+}$/H$^{+}$, has been derived
from the intensities of the [\NII]~$\lambda\lambda$\,6548,6584 lines. The
N/O abundance ratio was calculated assuming that N/O\,=\,N$^+$/O$^+$ and
N/H are estimated as log(N/H)\,=\,log(N/O)\,+\,log(O/H).

\item[(iv)] The neon ionic abundance, Ne$^{2+}$, has been derived using the
[\NeIII]\,$\lambda$\,3868\,\AA\ emission line, which is the only visible and
uncontaminated emission line of this atom in the spectra. For this ion we have
considered the electron temperature of [\OIII], as representative of the
high-ionization zone 
\citep[T$_e$({[}\NeIII{]})\,$\approx$\,T$_e$({[}\OIII{]});][]{1969BOTT....5....3P}.
\cite{2004A&A...415...87I} point out that this assumption can lead to an
overestimate of Ne/H in objects with low excitation, where the charge transfer
between O$^{2+}$ and H$^0$ becomes important. Thus, we have used the
expression of this ICF given by \cite{2007MNRAS.381..125P}. Given the high
excitation of the observed objects, there is no significant differences
between the derived neon abundances using this ICF and those estimated with
the classical approximation (H08, H11).

\item[(v)]The argon ionic abundance, Ar$^{2+}$, has been calculated from the
measured [\ArIII]\,$\lambda$\,7136\,\AA\ emission line assuming that
T$_e$([\ArIII])~$\approx$\,T$_e$([\SIII]) \citep{1992AJ....103.1330G}. In some
cases, it is possible to measure the [\ArIV]\,$\lambda\lambda$\,4711,4740\,\AA\ emission lines. However, the first one is usually merged with the
\HeI\,$\lambda$\,4713\,\AA\ line, and we are not able to separate
them. Therefore, we use the second and more intense line to calculate the
Ar$^{3+}$ abundance. The ionic abundance of Ar$^{3+}$ has been calculated
assuming that T$_e$([\ArIV])\,$\approx$\,T$_e$([\OIII]). The total abundance
of argon are 
hence calculated using the ICF(Ar$^{2+}$) and the ICF(Ar$^{2+}$\,+\,Ar$^{3+}$)
derived from photoionization models by \cite{2007MNRAS.381..125P}. 

\item[(vi)]Finally, for iron we have used the [\FeIII]\,$\lambda$\,4658\,\AA\
emission line using the electron temperature of [\OIII]
[T$_e$([\FeIII])\,$\approx$\,T$_e$([\OIII])]. We have taken the ICF(Fe$^{2+}$)
from \cite{2004IAUS..217..188R}. We were only able to measure the
[\FeIII]$\lambda$\,4658\,\AA\ emission line flux in the long-slit spectrum of
knot B. 

\end{description}

The ionic abundances with respect to ionized hydrogen of the elements heavier
than helium, ICFs, total abundances and their corresponding errors are also
given in Tables \ref{abundancesBC} and \ref{abundancesABCEF} for long-slit and
echelle data, respectively. 

\begin{table}
{\scriptsize
\caption[]{Physical properties, and ionic and total chemical abundances
  derived from forbidden lines and the helium recombination lines for the
  global measurements of the long-slit data: knots B and C.} 
\label{abundancesBC}
\begin{center}
\begin{tabular}{lcc}
\hline
 &B&C\\
\hline
  n([\SII]) &100: & 100:         \\
  t$_e$([\OIII]) &        1.26$\pm$0.03    & 1.01$\pm$0.02${^c}$    \\
  t$_e$([\SIII]) & 1.19$\pm$0.12 & 0.88$\pm$0.13${^d}$${^e}$\\
  t$_e$([\OII]) & 1.20$\pm$0.04 & 1.21$\pm$0.01${^a}$\\
  t$_e$([\SII]) &  0.79$\pm$0.11  &1.21$\pm$0.01${^a}$\\ 
  t$_e$([\NII]) & 1.20$\pm$0.04${^a}$  &0.93$\pm$0.01${^e}$\\ [4pt]

12+$\log(O^+/H^+)$        &  7.34$\pm$0.05 &  7.82$\pm$0.06${^a}$
  \\
 12+$\log(O^{2+}/H^+)$     &  7.95$\pm$0.03 &  7.84$\pm$0.03${^c}$ 
  \\
 \bf{12+log(O/H)}          &  8.04$\pm$0.03 &  8.13$\pm$0.04${^b}$
 \\[2pt]
  
 12+$\log(S^+/H^+)$        &  5.95$\pm$0.20 &  6.16$\pm$0.07${^a}$
  \\
 12+$\log(S^{2+}/H^+)$     &  6.17$\pm$0.23 &  \ldots
  \\
 ICF($S^++S^{2+}$)         &  1.41$\pm$0.03 &   \ldots
  \\
 \bf{12+log(S/H)}          &  6.52$\pm$0.22 &   \ldots
  \\
 \bf{log(S/O)}           & -1.52$\pm$0.22 &  \ldots
  \\[2pt]
  
 12+$\log(N^+/H^+)$       &  6.18$\pm$0.04${^a}$ &  6.70$\pm$0.07${^e}$
  \\
   \bf{12+log(N/H)}             & 6.88$\pm$0.20${^a}$ & 7.02$\pm$0.23${^e}$
  \\
 \bf{log(N/O)}             & -1.16$\pm$0.07${^a}$ & -1.11$\pm$0.09${^e}$
 \\[2pt]
  
 12+$\log(Ne^{2+}/H^+)$    &  7.33$\pm$0.03 &  \ldots
  \\
 ICF($Ne^{2+}$)          &  1.08$\pm$0.01 &  \ldots
  \\
 \bf{12+log(Ne/H)}         &  7.36$\pm$0.03 &  \ldots
  \\
 \bf{log(Ne/O)}           & -0.68$\pm$0.04 &  \ldots
  \\[2pt]
  
 12+$\log(Ar^{2+}/H^+)$    &  5.65$\pm$0.10 &  \ldots
  \\
 12+$\log(Ar^{3+}/H^+)$    &  4.62$\pm$0.08 &  \ldots
  \\
 ICF($Ar^{2+}$+$Ar^{3+}$)  &  1.03$\pm$0.01 &  \ldots
  \\
 \bf{12+log(Ar/H)}        &  5.70$\pm$0.10 &  \ldots
 \\
 \bf{log(Ar/O)}            & -2.34$\pm$0.11 &\ldots
 \\ [2pt]
  
 12+$\log(Fe^{2+}/H^+)$    &  5.04$\pm$0.10 &  \ldots
  \\
 ICF($Fe^{2+}$)            &  5.21$\pm$0.62 &  \ldots
  \\
 \bf{12+log(Fe/H)}         &  5.76$\pm$0.11 &  \ldots
  \\[2pt]
  
 $He^+/H^+$($\lambda$\,4471)    & 0.064$\pm$0.003 & \ldots
  \\
  $He^+/H^+$($\lambda$\,5876)    & 0.096$\pm$0.002 & \ldots
  \\
   $He^+/H^+$($\lambda$\,6678)    & 0.072$\pm$0.003 & \ldots
  \\
 $He^+/H^+$($\lambda$\,7065)    & 0.100$\pm$0.008 & \ldots
  \\
  $He^+/H^+$adopted    & 0.086$\pm$0.015 & \ldots
  \\
$He^{2+}/H^+$($\lambda$\,4686)    & 0.0013$\pm$0.0002 & \ldots
  \\
\bf{(He/H)}                & 0.087$\pm$0.015 & \ldots
  \\[2pt]
  \hline
\noalign{\smallskip}
\multicolumn{3}{@{}p{0.4\textwidth}@{}}{\footnotesize {Note. All the values with a
    superscript are derived using models or empirical/semi-empirical
    relations, and the derived errors are underestimated.} Densities in $cm^{-3}$ and temperatures in 10$^4$\,K}\\
\multicolumn{3}{@{}p{0.4\textwidth}@{}}{\footnotesize  {${^a}$ derived using
  temperatures predicted by photoionization models, see text. ${^b}$derived using
  temperatures estimated from photoionization models and/or empirical
  methods. ${^c}$derived using H06 empirical method. ${^d}$derived
  using D07 empirical method. ${^e}$derived using P07 empirical method.}}
\end{tabular}
\end{center}}
\end{table}

\landscape
\begin{table} 
\vspace*{3cm}
 {\scriptsize
\caption[]{{\footnotesize Physical properties, and ionic and total chemical
    abundances derived from forbidden lines and the helium recombination lines 
    for the global and the different kinematic measurements in the
    echelle data: knots A, B, C and E.}}
\label{abundancesABCEF}
\begin{center}
\begin{tabular}{@{}l@{\hspace{0.2cm}}c@{\hspace{0.2cm}}c@{\hspace{0.2cm}}c@{\hspace{0.2cm}}c@{\hspace{0.3cm}}c@{\hspace{0.2cm}}c@{\hspace{0.2cm}}c@{\hspace{0.3cm}}c@{\hspace{0.2cm}}c@{\hspace{0.2cm}}c@{\hspace{0.2cm}}c@{\hspace{0.2cm}}c@{\hspace{0.2cm}}c@{}}
\hline
 &\multicolumn{4}{@{\hspace{0.2cm}}c}{A}&\multicolumn{3}{c}{B}&\multicolumn{3}{c}{C}&\multicolumn{3}{c@{}}{E}\\
 & global&  narrow\,1 & narrow\,2  & broad &global &narrow &broad& global &narrow &broad &global &narrow\,1 &narrow\,2\ \\
 \hline
 n([\SII])  &300: & 100:   & 200: & 100:    &150$\pm$50 & 100:   & 150:  &400: & 400:   & 1400: &500: & 900:   & 400:           \\     
 t$_e$([\OIII]) &1.00$\pm$0.02${^c}$    & 1.08$\pm$0.03${^c}$&     1.16$\pm$0.04${^c}$  & 0.95$\pm$0.05${^c}$  &1.20$\pm$0.06 &1.22$\pm$0.18 & 1.12$\pm$0.04& 1.0$^f$ &1.0$^f$&1.0$^f$& 0.97$\pm$0.03${^c}$  & 1.0$^f$  &   1.0$^f$\\
 t$_e$([\SIII])  & 0.87$\pm$0.05${^d}$    &0.96$\pm$0.06${^d}$&    1.06$\pm$0.10${^d}$    & 0.8$\pm$0.12${^d}$ &1.31$\pm$0.17 & 1.71$\pm$0.30&1.16$\pm$0.16 & 0.87$\pm$0.13${^a}$ &0.87$\pm$0.13${^a}$ & 0.87$\pm$0.13${^a}$ & 0.83$\pm$0.13${^d}$${^e}$& 0.87$\pm$0.13${^a}$   & 0.87$\pm$0.13${^a}$\\
 t$_e$([\OII])${^a}$  &0.93$\pm$0.01 & 1.28$\pm$0.01& 1.06$\pm$0.01& 1.08$\pm$0.03& 1.11$\pm$0.02 &1.28$\pm$0.10&      1.07$\pm$0.01  &  0.87$\pm$0.01 &0.87$\pm$0.01 & 0.76$\pm$0.01  &   0.84$\pm$0.01       & 0.80$\pm$0.01  &  0.88$\pm$0.01\\
  t$_e$([\SII])${^a}$ &0.93$\pm$0.01 & 1.28$\pm$0.01& 1.06$\pm$0.01&1.08$\pm$0.03&   1.11$\pm$0.02  &1.28$\pm$0.10&       1.07$\pm$0.01& 0.87$\pm$0.01 & 0.87$\pm$0.01 & 0.76$\pm$0.01   &  0.84$\pm$0.01       & 0.80$\pm$0.01  &  0.88$\pm$0.01 \\
 t$_e$([\NII])  & 0.93$\pm$0.01${^a}$& 1.28$\pm$0.01${^a}$ &  1.06$\pm$0.01${^a}$&  1.08$\pm$0.03${^a}$& 1.11$\pm$0.02${^a}$  &  1.28$\pm$0.10${^a}$&    1.07$\pm$0.01${^a}$& 0.87$\pm$0.01${^a}$  & 0.87$\pm$0.01${^a}$  & 0.76$\pm$0.01${^a}$     & 0.88$\pm$0.01${^e}$   & 0.80$\pm$0.01${^a}$   & 0.88$\pm$0.01${^a}$\\ [2pt]
 
 12+$\log(O^+/H^+)$${^a}$          &  7.92$\pm$0.06 &  7.87$\pm$0.05 &  7.87$\pm$0.10 &  8.00$\pm$0.15 &  7.21$\pm$0.08 &  7.61$\pm$0.21 &  7.09$\pm$0.08 &  \ldots &  \ldots &  \ldots &  8.01$\pm$0.19 &  \ldots &  \ldots \\
  12+$\log(O^{2+}/H^+)$     &  7.80$\pm$0.03${^c}$ &  7.55$\pm$0.04${^c}$ &  7.75$\pm$0.05${^c}$ &  8.01$\pm$0.11${^c}$ &  8.11$\pm$0.07 &  7.84$\pm$0.19 &  8.26$\pm$0.05 &  7.83$\pm$0.03${^a}$ &  7.99$\pm$0.02${^a}$ &  7.71$\pm$0.05${^a}$ &  7.70$\pm$0.08${^c}$ &  7.53$\pm$0.04$^f$ &  7.84$\pm$0.02$^f$ \\
 \bf{12+log(O/H)}${^b}$          &  8.17$\pm$0.05 &  8.04$\pm$0.05 &  8.11$\pm$0.08 &  8.31$\pm$0.13 &  8.16$\pm$0.07 &  8.04$\pm$0.20 &  8.29$\pm$0.05 &  \ldots &  \ldots &  \ldots &  8.18$\pm$0.16${^b}$ &  \ldots &  \ldots  \\[2pt]
  
 12+$\log(S^+/H^+)$${^a}$        &  6.14$\pm$0.04 &  5.90$\pm$0.04 &  5.71$\pm$0.07 &  6.44$\pm$0.09 &  5.30$\pm$0.05 &  5.68$\pm$0.12 &  5.02$\pm$0.08 &  6.20$\pm$0.17 &  6.24$\pm$0.16 &  6.37$\pm$0.28 &  6.36$\pm$0.17 &  5.99$\pm$0.20 &  6.56$\pm$0.16  \\
 12+$\log(S^{2+}/H^+)$     &  6.42$\pm$0.18${^d}$ &  6.11$\pm$0.18${^d}$ &  6.20$\pm$0.16${^d}$ &  6.54$\pm$0.28${^d}$ &  5.93$\pm$0.12 &  5.79$\pm$0.13 &  6.03$\pm$0.13 &  \ldots &  \ldots &  \ldots &  \ldots &  \ldots &  \ldots\\
 ICF($S^++S^{2+}$)${^b}$    &  1.05$\pm$0.01 &  1.02$\pm$0.01 &  1.05$\pm$0.01 &  1.08$\pm$0.02 &  1.72$\pm$0.02 &  1.16$\pm$0.01 &  2.13$\pm$0.07 & \ldots & \ldots & \ldots &  \ldots & \ldots & \ldots  \\
 \bf{12+log(S/H)}${^b}$    &  6.62$\pm$0.13 &  6.33$\pm$0.13 &  6.34$\pm$0.14 &  6.83$\pm$0.21 &  6.26$\pm$0.11 &  6.11$\pm$0.13 &  6.40$\pm$0.12 & \ldots &  \ldots &  \ldots &  \ldots &  \ldots &  \ldots \\
 \bf{log(S/O)}${^b}$       & -1.54$\pm$0.14 & -1.71$\pm$0.14 & -1.77$\pm$0.16 & -1.48$\pm$0.25 & -1.90$\pm$0.13 & -1.93$\pm$0.23 & -1.88$\pm$0.13 &  \ldots &  \ldots &  \ldots & \ldots &  \ldots &  \ldots \\[2pt]
  
 12+$\log(N^+/H^+)$        &  6.97$\pm$0.03${^a}$&  6.89$\pm$0.03${^a}$ &  6.78$\pm$0.05${^a}$ &  7.06$\pm$0.10${^a}$ &  5.87$\pm$0.05${^a}$ &  6.25$\pm$0.13${^a}$ &  5.75$\pm$0.06${^a}$ &  6.57$\pm$0.15${^a}$ &  6.78$\pm$0.12${^a}$ &  6.35$\pm$0.19${^a}$ &  6.92$\pm$0.12${^e}$ &  6.65$\pm$0.11${^a}$&  7.15$\pm$0.12${^a}$ \\
   \bf{12+log(N/H)}             & 7.21$\pm$0.23${^a}$ &  7.06$\pm$0.22${^a}$ & 7.06$\pm$0.22${^a}$ & 7.02$\pm$0.39${^a}$ & 6.82$\pm$0.32${^a}$  &  6.68:${^a}$ & 6.95$\pm$0.28${^a}$ &  \ldots &  \ldots &  \ldots & 7.09:${^e}$ &  \ldots &  \ldots \\
 \bf{log(N/O)}             & -0.96$\pm$0.06${^a}$ & -0.98$\pm$0.06${^a}$ & -1.09$\pm$0.11${^a}$ & -0.94$\pm$0.18${^a}$ & -1.34$\pm$0.09${^a}$ & -1.36$\pm$0.25${^a}$ & -1.34$\pm$0.10${^a}$ & \ldots &  \ldots &  \ldots & -1.09$\pm$0.23${^e}$ & \ldots &  \ldots \\[2pt]
  
 12+$\log(Ne^{2+}/H^+)$    &  7.50$\pm$0.07${^c}$ &  \ldots &  \ldots &  \ldots &  7.49$\pm$0.09 &  7.34$\pm$0.28 &  7.62$\pm$0.07 &  \ldots &  \ldots &  \ldots &  \ldots &  \ldots &  \ldots \\
 ICF($Ne^{2+}$)            &  1.21$\pm$0.01${^c}$ &  \ldots &  \ldots &  \ldots &  1.07$\pm$0.01 &  1.11$\pm$0.01 &  1.07$\pm$0.01 &  \ldots &  \ldots &  \ldots &  \ldots &  \ldots &  \ldots \\
 \bf{12+log(Ne/H)}         &  7.59$\pm$0.07${^c}$ &  \ldots &  \ldots &  \ldots &  7.52$\pm$0.09 &  7.39$\pm$0.28 &  7.65$\pm$0.07 &  \ldots &  \ldots &  \ldots &  \ldots &  \ldots &  \ldots \\
 \bf{log(Ne/O)}            & -0.58$\pm$0.09${^c}$ &  \ldots &  \ldots &  \ldots & -0.64$\pm$0.11 & -0.65$\pm$0.35 & -0.64$\pm$0.09 &  \ldots &  \ldots &  \ldots &  \ldots &  \ldots &  \ldots \\[2pt]
  
 12+$\log(Ar^{2+}/H^+)$    &  6.05$\pm$0.19${^d}$ &  5.76$\pm$0.20${^d}$ &  5.63$\pm$0.20${^d}$ &  6.36$\pm$0.31${^d}$ &  5.56$\pm$0.11 &  5.50$\pm$0.14 &  5.65$\pm$0.14 &  \ldots &  \ldots &  \ldots &  \ldots &  \ldots &  \ldots \\
 ICF($Ar^{2+}$)            &  1.15$\pm$0.01${^d}$ &  1.19$\pm$0.01${^d}$ &  1.15$\pm$0.01${^d}$ &  1.13$\pm$0.01${^d}$ &  1.38$\pm$0.02 &  1.11$\pm$0.01 &  1.79$\pm$0.08 &  \ldots &  \ldots &  \ldots &  \ldots &  \ldots &  \ldots \\
 \bf{12+log(Ar/H)}         &  6.11$\pm$0.19${^d}$ &  5.83$\pm$0.20${^d}$ &  5.69$\pm$0.20${^d}$&  6.41$\pm$0.31${^d}$ &  5.70$\pm$0.11 &  5.55$\pm$0.14 &  5.90$\pm$0.14 &  \ldots &  \ldots &  \ldots &  \ldots &  \ldots &  \ldots \\
  \bf{log(Ar/O)}            & -2.06$\pm$0.20${^d}$ & -2.21$\pm$0.21${^d}$ & -2.43$\pm$0.21${^d}$ & -1.90$\pm$0.34${^d}$ & -2.46$\pm$0.13 & -2.49$\pm$0.24 & -2.39$\pm$0.15 &  \ldots &  \ldots &  \ldots &  \ldots &  \ldots &  \ldots \\  [2pt]

$He^+/H^+$($\lambda$\,4471)    & 0.094$\pm$0.018${^a}$ & \ldots & \ldots & \ldots & 0.096$\pm$0.003 & \ldots & \ldots & \ldots & \ldots & \ldots & \ldots & \ldots & \ldots \\
 $He^+/H^+$($\lambda$\,5876)    & 0.072$\pm$0.004${^a}$ & 0.051$\pm$0.005${^a}$ & 0.101$\pm$0.010${^a}$ & 0.111$\pm$0.024${^a}$ & 0.079$\pm$0.001 & 0.044$\pm$0.001 & 0.092$\pm$0.001 & \ldots & \ldots & \ldots & 0.124$\pm$0.040${^a}$ & \ldots & \ldots \\
 $He^+/H^+$($\lambda$\,6678)    & 0.109$\pm$0.012${^a}$ & \ldots & \ldots & \ldots & 0.086$\pm$0.001 & 0.035$\pm$0.006 & 0.113$\pm$0.003 & \ldots & \ldots & \ldots & \ldots & \ldots & \ldots \\
 $He^+/H^+$($\lambda$\,7065)    & 0.107$\pm$0.036${^a}$ & \ldots & \ldots & \ldots & 0.109$\pm$0.005 & \ldots & \ldots & \ldots & \ldots & \ldots & \ldots & \ldots & \ldots \\
 $He^+/H^+$adopted    & 0.077$\pm$0.023${^a}$ & 0.051$\pm$0.005${^a}$ & 0.101$\pm$0.010${^a}$ & 0.111$\pm$0.024${^a}$ & 0.080$\pm$0.017 & 0.043$\pm$0.006 & 0.094$\pm$0.014 & \ldots & \ldots & \ldots & 0.124$\pm$0.040${^a}$ & \ldots & \ldots 
\\[2pt]
\hline
\noalign{\smallskip}
\multicolumn{14}{@{}p{1.3\textwidth}@{}}{\footnotesize {Note. All the values with a
    superscript are derived using models or empirical/semi-empirical
    relations, and the derived errors are underestimated.} Densities in
  $cm^{-3}$ and temperatures in 10$^4$\,K}\\ 
\multicolumn{14}{@{}p{1.3\textwidth}@{}}{\footnotesize {${^a}$derived using
  temperatures predicted by photoionization models, see text. ${^b}$derived using
  temperatures estimated from photoionization models and/or empirical
  methods. ${^c}$derived using H06 empirical method. ${^d}$derived using D07
  empirical method. ${^e}$derived using P07 empirical method. $^f$assumed
  temperature\,=\,10$^4$\,K}} 
\end{tabular}
\end{center}}
\end{table}
\endlandscape

\section{Discussion}
\label{sec:Discus}
In the following subsections we will discuss the results obtained from the two different observation modes. 

\subsection{Long-slit data}
\label{sec:Long-slitdata} 

\subsubsection*{\Haro\,B}

Four electron temperatures, T$_e$([\OIII]), T$_e$([\OII]), T$_e$([\SIII]) and
T$_e$([\SII]), were estimated from direct measurements for knot B. The
good quality of the data translates into small rms errors (2\%, 3\%, 10\% and
14\%, respectively, for each temperature). In the absence of the
[\NII]~$\lambda$\,5755\AA\ emission line we assumed the 
relationships between temperatures to be T$_e$([\NII])~$\approx$\,T$_e$([\OII]).

Knot B shows an electron density value of n$_e$\,=\,100 cm$^{-3}$, typical for
this kind of objects. However, the error is too large, thus it should be
regarded as an estimation of the order magnitude of the density. 

In this knot, we have derived the ionic abundances of O$^{+}$, O$^{2+}$,
S$^{+}$, S$^{2+}$, N$^{+}$, Ne$^{2+}$, Ar$^{2+}$, Ar$^{3+}$ and  Fe$^{2+}$
using the direct method as described in H08.
We have also calculated the total
abundances of O, S, N, Ne, Ar and Fe, and the logarithmic ratios of N/O, S/O,
Ne/O and Ar/O.

The oxygen abundance is 8.04$\pm$0.03, which is 0.22 times the solar value
\cite[12+log(O/H)\,$_\odot$\,=\,8.69;][]{2001ApJ...556L..63A}. This oxygen
abundance value is in very good agreement, within the observational errors,
to that found by \cite{2009A&A...508..615L}, 8.10$\pm$0.06. The sulphur
abundance is 0.15 the solar value
\cite[12+log(S/H)\,$_\odot$\,=\,7.33;][]{1998SSRv...85..161G}, and the
nitrogen abundance is 0.09 of the solar one
\cite[12+log(N/H)\,$_\odot$\,=\,7.93;][]{2001AIPC..598...23H}.

\begin{figure*}
\centering
\includegraphics[trim=0cm 0cm 0cm 0cm,clip,angle=0,width=8.5cm,height=8.5cm]{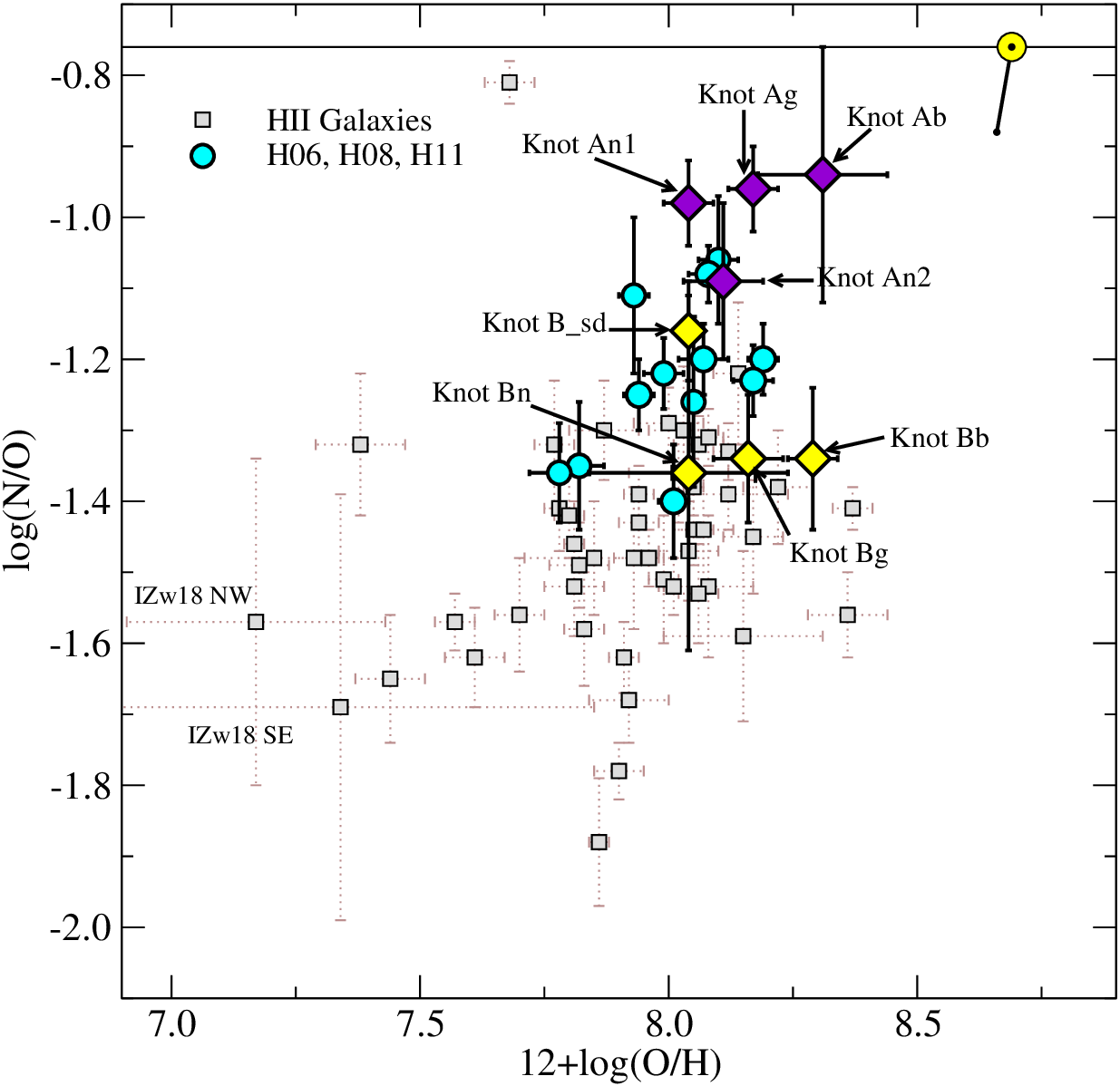}\hspace{0.3cm}
\includegraphics[trim=0cm 0cm 0cm 0cm,clip,angle=0,width=8.5cm,height=8.5cm]{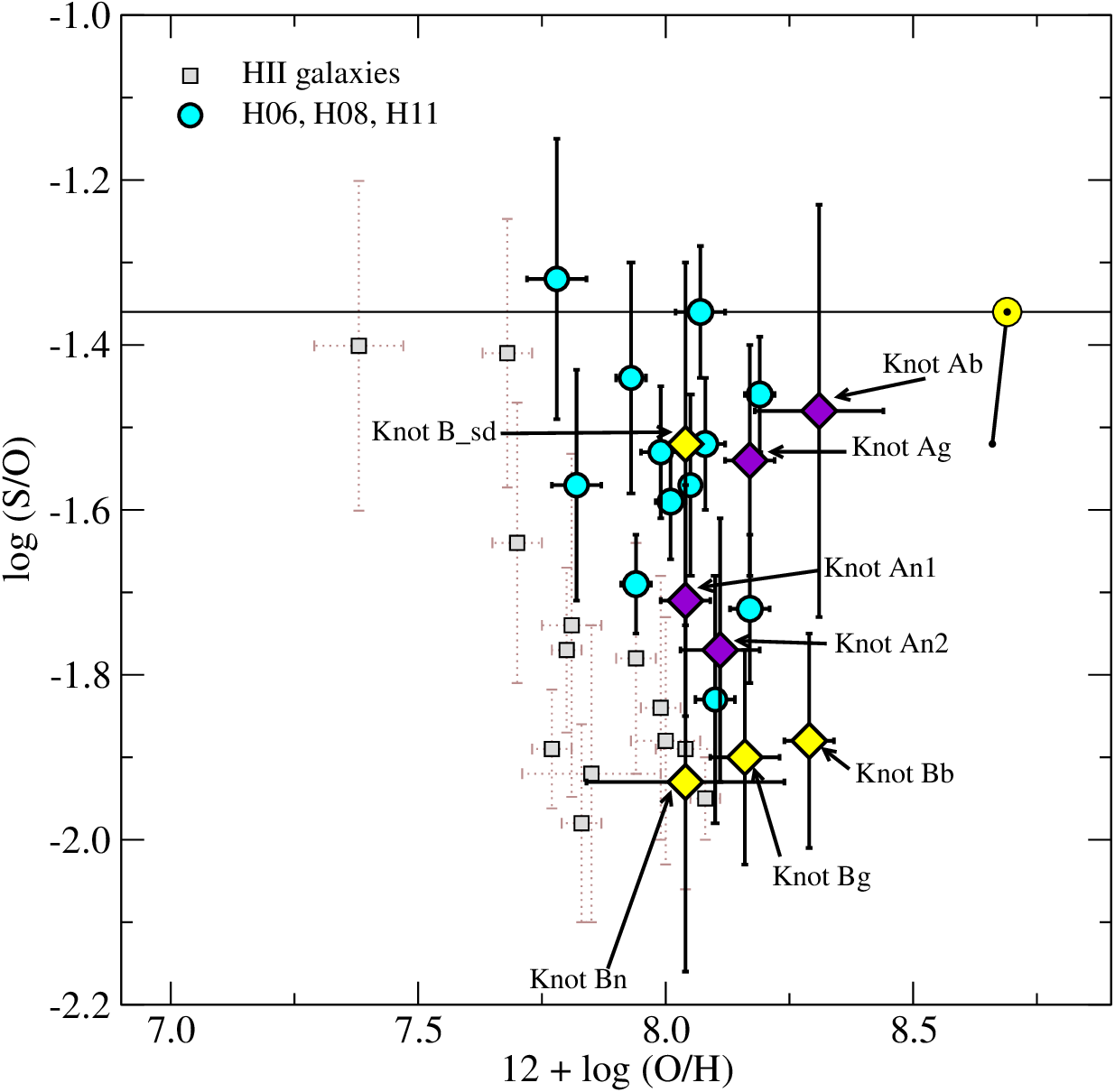}
\caption[]{{N/O and S/O as a function of 12+log(O/H) (left and right panel, respectively) for knot A and B
(open and filled diamonds, violet and yellow, respectively), and the objects from H06, H08, and H11 (open circles,
turquoise). Open squares are the \HII\ galaxies from Table 13
of H08.} The
observed knots are marked in the plot with their names plus a letter which
denote the observation mode (``sd'': single dispersion) or the kinematic
component measure (``g'': global, ``n'': narrow, and ``b'': broad). The
solar values are shown with the usual sun symbol: oxygen from
\citet{2001ApJ...556L..63A}, nitrogen from \citet{2001AIPC..598...23H} and
sulphur from \citet{1998SSRv...85..161G}. These values are linked by a solid line with the
solar ratios from \citet{2005ASPC..336...25A}.} 
\label{fig:abrelativas}
\end{figure*}

The logarithmic N/O ratio found for knot B is 0.28\,dex lower than the solar
value \cite[log(N/O)\,$\odot$\,=\,-0.88;][]{2005ASPC..336...25A}, and 0.44\,dex
higher than the typical value shown by this kind of
objects which present a plateau at log(N/O) about
-1.6\,dex \citep{1979MNRAS.189...95P,1979A&A....78..200A} in the low
metallicity end of the distribution \cite[see discussion in][and 
references therein]{2010ApJ...715L.128A,Perez-Montero+11}. The derived value
is on the upper range of the distribution of the log(N/O) ratio for this kind
of objects (see Fig.\ \ref{fig:abrelativas}).
The log(S/O) ratio is -1.52$\pm$0.22, in agreement, within the errors, with
the solar value \cite[log(S/O)\,$_\odot$\,=\,-1.36;][]{1998SSRv...85..161G}
(see Fig.\ \ref{fig:abrelativas}). 
The logarithmic Ne/O and Ar/O ratios are also consistent with the solar ones
\cite[log(Ne/O)\,$_\odot$\,=\,-0.61\,dex and
log(Ar/O)\,$_\odot$\,=\,-2.29\,dex;][]{1998SSRv...85..161G} within the errors
{(see Tables \ref{abundancesBC} and \ref{abundancesABCEF})}.

We have used the intensities of the well detected \HeI~$\lambda\lambda$\,4471,5876,6678 and
7065\AA, and \HeII\,$\lambda$\,4686\,\AA\ lines to estimated the ionic abundances of 
He$^{+}$ and He$^{2+}$, respectively. Then we have derived the helium total
abundance using these ionic abundances. The adopted value for He$^{+}$/H$^{+}$
is 0.086$\pm$0.015 and 0.0013$\pm$0.0002 for He$^{2+}$/H$^{+}$. The He total
abundance estimated in this region is 0.087$\pm$0.015, within the typical
values found for \HII\ galaxies (see e.g.\ H08, and references therein).


\subsubsection*{\Haro\,C}

In the case of knot C it was impossible to obtain any electron temperature using
the direct method. In this case, we estimated the temperatures using the
models and the empirical relationships, as explained in the previous section.

The auroral lines are not detected in the knot C spectrum due to the low S/N
of these data, so we have estimated a [\NII] temperature of
9300\,$\pm$\,100\,K 
following the Pilyugin's method (P07). Using the approximation
T$_e$([\NII])\,$\approx$\,500K\,+\,T$_e$([\SIII]) derived by D07, we have
estimated T$_e$([\SIII]), and derived T$_e$([\OIII]) from the
empirical relation fitted by H06. The {electron} temperature T$_e$([\OII]) has
been calculated from T$_e$([\OIII]) using the relationship based on
photoionization models derived by \cite{2003MNRAS.346..105P}, and from the
approximation T$_e$([\SII])\,$\approx$\,T$_e$([\OII]) we obtained the electron
temperature of [\SII]. As in the case of knot B, due to the large errors involved, the derived electron density of 100 particles per cm$^3$ has to be regarded as a rough estimation.

In this knot, we were not able to measure the [\FeIII]~$\lambda$\,4658 emission line and therefore no iron abundance could be derived. The [\NII]\,$\lambda$\,6548\,\AA\ emission line could not be measured either and was taken to be 0.33 times that of [\NII]\,$\lambda$\,6584\,\AA.


\subsubsection*{Comparison among knots}

{A comparison between the results for both knots from long-slit data might suggest, at face
value, that the oxygen abundance could be slightly higher in knot C. However,
the 1\,$\sigma$ errors derived for these oxygen abundances do not take into
account the errors in the line temperatures introduced by uncertainties in the  
calibrations of the relations based on photoionization models. These
unaccounted effect will most probably make the observed difference negligible.
The estimated N/O ratio for knot C is only marginally larger than that for knot B,
but still the same within the observational errors. 

The temperatures representative of the high-ionization zone are higher for knot B 
than for knot C. The temperatures for the low-ionization zone
show a different behaviour: in the case of T$_e$([\OII]) appear to be similar
in both knots while the T$_e$([\SII]) in knot B is 4200\,K lower than in knot
C. We have to note that the [\SII] temperature was derived using the direct
method for knot B, but using photoionization models for knot C. 
This leads to a difference in temperature among [\OII] and [\SII] for knot B,
although models 
state that we should expect similar values.
This behaviour was previously found for several
objects \cite[see Fig.\ 2.17 of][]{2008PhDT........35H}. Moreover,
\cite{2010MNRAS.404.2037P} presented a self-consistent study in a sample of 10
\HII\ galaxies computing tailored models with the photoionization code CLOUDY
\citep{Ferland+98}. They found that the 
electron temperature of [\SII] was overestimated by the models, with the
corresponding underestimate of its abundance, pointing to the possible
presence of outer shells of diffuse gas in these objects that were not
taken into account in their models.
In contrast, the [\NII] temperature shows for knot B a higher value than that
derived for knot C by about 2700\,K, but again they were estimated using two
different methods. Only for comparison purposes we have also estimated this
temperature for knot B using Pilyugin's method finding
T$_e$([\NII])\,=\,10500\,$\pm$\,100\,K, which is lower than the estimated
using the models by 1500\,K. Comparing those [\NII] temperatures derived using
Pilyugin's method, knot B still shows higher values than knot C by about 
1200\,K.}

The excitation degree, given by the logarithmic O$^{2+}$/O$^{+}$ ratio, is
higher in knot B (0.61\,dex) than in knot C (0.02\,dex), in agreement with the
estimated values by \cite{2009Ap&SS.324..355L}. The estimated extinction in
both knots is small, typical
of this kind of objects \cite[][and references therein]{2008PhDT........35H}.

\subsection{Echelle data}
\label{sec:Echelle data} 
We analysed the parameters derived from the global and the different
kinematical component measurements for the four knots using the echelle data. 

\subsubsection*{\Haro\,A}

A large number of recombination and forbidden lines were detected in our
echelle spectrum, although we were not able to measure any auroral emission
line. We have therefore used the relation between the SO$_{23}$ parameter and
the [\SIII] temperature derived by D07 to estimate T$_e$([\SIII]). From
this temperature we have calculated T$_e$[\OIII] using the relationship found
by H06. Following the classical analysis (see H08), we estimated
T$_e$([\OII]), T$_e$([\SII]) and T$_e$([\NII]) using the relations given by
photoionization models. 

{The estimated 
density errors are large, hence these values are only considered as an order
of magnitude. All the electron density determinations are well below the
critical value for collisional de-excitation.}

The global measure present a total oxygen abundance of 8.17\,$\pm$\,0.05, 0.3
times the solar value, while for the different kinematical components we
obtained 8.04\,$\pm$\,0.05 (narrow\,1), 8.11\,$\pm$\,0.08 (narrow\,2), and
8.31\,$\pm$\,0.13 (broad), 0.22, 0.26 and 0.42 times the solar abundance,
respectively. 
There seems to be a difference between broad and narrow components above 
1\,$\sigma$ level, but since the errors in the calibration of the relations based on
photoionization models are not quantitatively established, we can
consider the total abundances in agreement for all components.
 The derived total oxygen abundance for the different
kinematical components and for the global measure in this knot present the
characteristic low values, within the errors, that are found for \HII\ galaxies: 12+log(O/H) between 7.94 and 8.19 
\citep{1991A&AS...91..285T,2006MNRAS.365..454H}. 

{Taking into account the errors,
the total sulphur abundance of the broad component is about 0.15\,dex higher than
those shown by the narrow ones, but in agreement considering errors at
2\,$\sigma$ level. On the other hand, the values for total nitrogen abundances
are in very good agreement for all the components.}

The ionic abundances of sulphur, S$^{+}$/H$^{+}$ and S$^{2+}$/H$^{+}$, the
ionic nitrogen abundance, N$^{+}$/H$^{+}$, and the ionic argon abundance,
Ar$^{2+}$/H$^{+}$, obtained for the broad component are higher than the ionic
abundances found for the narrow components (between 0.17 to 0.61\,dex respect
to the narrow\,1 component, and between 0.28 to 0.73\,dex respect to the narrow
2 one). Only for the global measure it was possible to obtain the ionic and total
abundances of [\NeIII] from the measure of the neon 3868\,\AA\ emission
line. 

{We have also derived the sulphur and nitrogen abundances relative to oxygen
abundance, S/O and N/O. 
If we consider the errors, all the S/O derived values
are very similar among them (see Fig. \ref{fig:abrelativas}).} The N/O value
estimated for the narrow\,2 component is the lowest, but its difference with
the other components is negligible. This behaviour of the N/O could be an
evidence of a common or very similar chemical evolution for the different
kinematical components \cite[see discussion in][]{Perez-Montero+11}. It could be
also indicative that the different kinematical components correspond to different phases
of the same gas, which is in agreement with the scenarios proposed, for
example, by \cite{Tenorio-Tagle+96} or \cite{2009ApJ...706.1571W}. In knot A
we also obtained an excess of the N/O ratio (see Fig. \ref{fig:abrelativas})
respect to the typical values found for an \HII\ galaxy. 

{For the global measure we can determine the ionic helium abundance,
from the 4471, 5876, 6678, 7065\,\AA\ emission lines.
For the three kinematical components we were only able to deconvolve the 
5876\,\AA\ emission line.
The narrow\,1 component
has a lower ionic abundance than the typical values found for other similar
galaxies, while the values found for the other two components and the global
measure are in the range presented by \HII\ galaxies (see H08).}

\subsubsection*{\Haro\,B}

In this knot we were able to estimate T$_e$([\OIII]), and
T$_e$([\SIII]) using the direct method through the deconvolution of the
emission profile for the different kinematical components of
[\OIII]\,$\lambda$\,4363\,\AA\ and [\SIII]\,$\lambda$\,6312\,\AA\ auroral
emission lines. The precision obtained for T$_e$([\OIII])
and T$_e$([\SIII]) are of the order of 5\% and 13\% for the global measure,
15\% and 18\% for the narrow component, and  4\% and 14\% for the broad
component, respectively for each temperature. These are the only two
temperatures directly derived, even for the global measure. The [\OII]
temperature has been obtained from T$_e$([\OIII]) using the relations given by
photoionization models \citep{2003MNRAS.346..105P}. The [\SII] and
[\NII] temperatures have been estimated using the approximation
T$_e$([\SII])\,$\approx$\,T$_e$([\NII])\,$\approx$\,T$_e$([\OII]). 

{The errors in the electron densities are large enough as
to consider our estimate as an order of magnitude.} Once again, in
all cases the electron densities are well below the critical value for
collisional de-excitation.

The ionic abundances in the low-ionization zone (O$^{+}$/H$^+$, S$^{+}$/H$^+$,
N$^{+}$/H$^+$) derived for the narrow component are higher than those shown by
the broad one, while the ions in the medium- and high-ionization zones
(S$^{2+}$/H$^+$ and Ar$^{2+}$/H$^+$, and O$^{2+}$/H$^+$ and Ne$^{2+}$/H$^+$,
respectively) show the opposite behaviour, lower values for the narrow
component.

{The total oxygen abundance derived for the global measure and the kinematical
components are in good agreement taking into account the observational errors,
where the global measure is approximately equal to the average of the derived
values for the different kinematical components weighted by luminosity. 
The values of the total sulphur abundances derived for both
kinematical components are very similar, with the broad component presenting
slightly larger values than the narrow one.}
In the case of the total nitrogen abundance we
derived very similar values for the different components, although we have to
note that the calculated error for the narrow component is too large, and
therefore this is only a rough estimate of the nitrogen abundance. The total
abundances of Ne 
derived for the different components are in agreement within the errors, and
the Ar total abundance from the broad component is slightly larger (by about
0.07\,dex) than that derived for the narrow one. As we said above, the
temperatures estimated using the relationships based on photoionization models
only take into account the errors of the line intensity measurements, but
without assigning any errors to the relation parameters. We could therefore state
that all the components present total abundances in overall good agreement. 

The S, N, Ne, and Ar abundance relative to oxygen for the narrow and
broad components, present mainly the same values, even comparing these values
with those estimated for the global measure (see
Fig. \ref{fig:abrelativas}). As in the case of knot A, this 
could be indicative that both components have a common or very similar
chemical evolution and/or that both components represent two different phases
of the same gas. 

{From the global measure, we have derived the ionic abundance of once ionized
helium
using the same emission lines as for knot A.
The adopted abundance value is
in good agreement to what we find in the long-slit
spectrum for the same knot, and in the range found for other \HII\ galaxies
(see H08). We have been able to deconvolve the profiles of
\HeI\,$\lambda\lambda$\,5876 and 6678\,\AA\  emission lines finding very low
values for the narrow component, 
while the broad component presents values 
typical for this class of objects.}

\subsubsection*{\Haro\,C and \Haro\,E}

In the cases of the star-forming knots C and E, we were not able to detect any
auroral emission line. Owing to the low S/N of these spectra, other some
important and relatively strong emission lines, such as
[\OII]\,$\lambda\lambda$\,3727\,\AA, do not have enough intensity to be
measured, even for the global measure. For all the components of knot C and the
kinematical components of knot E, we were not able to measure the
emission lines needed to apply the relationships given by D07 and P07 to derive
the [\SIII] and the [\NII] line temperatures, respectively. We assumed
that T$_e$([\OIII]) is equal to 10$^4$K (as was discussed in \S
\ref{sec:PhCond}), then we estimated the [\OII] temperature in each case using
its relation with T$_e$([\OIII]) given by photoinization models
\citep{2003MNRAS.346..105P}. T$_e$([\SII]) and T$_e$([\NII]) were
assumed equal to the [\OII] temperature, and T$_e$([\SIII]) was determined
using its relationship with the temperature of [\OIII] found by H06. 
For the global measure of knot E, it was possible to measure the
[\OII]\,$\lambda\lambda$\,3727,29\,\AA\ emission lines. Using these
emission lines together with the strong [\OIII] lines, we applied the
Pilyugin's method to estimate the electron temperature of [\NII]. Using
the systematic difference of 500\,K between the temperatures of [\NII] and
[\SIII] found by D07, we estimated T$_e$([\SIII]). From the [\SIII]
temperature we derived the T$_e$([\OIII]) using the relation by H06, then we
have used the same relations than for the kinematical components of knot E to
derive T$_e$([\OII]) and T$_e$([\SII]).


{The estimated electron densities for these two star-forming knots are rather
higher than those found for knots A and B.
Due to the very large uncertainties present in all the cases, these density
values are only estimates of the order of magnitude of the electron densities.
Only for the global measure of knot E,
it was possible to obtain an estimate of the ionic abundance of O$^{+}$ and,
thus, the total abundances of oxygen and nitrogen, and the N/O ratio. 
}

{In contrast with the behaviour found for the O$^{2+}$/H$^+$ ionic abundance for
the star-forming knots A and B, the ionic abundance of the narrow component
for knot C is higher ($\sim$\,0.3\,dex) than the one derived for the broad
component.  
This effect could be due to a different ionization degree of the kinematical
components.
In the case of knot E, 
the total oxygen abundance for the global measure is similar to the
abundances derived for knots A and B, and 
the helium abundance is slightly larger than that found for the other knots.
}


\subsubsection*{Comparison among knots}

The total oxygen abundance derived for the global measure of knot A is
slightly lower 
than the value obtained by L\'opez-S\'anchez \& Esteban
(\citeyear{2009A&A...508..615L}, 8.37\,$\pm$\,0.10). It could be due to the 
different hypotheses considered to determine the electron temperatures and the
chemical abundances. The total oxygen abundance estimated from the global
measure for knot B is in very good agreement with the value obtained by
L\'opez-S\'anchez \& Esteban (8.10\,$\pm$\,0.06) for the same knot. 
{We have estimated its electron temperature of [\OIII] 
using the direct method, and it is in very good agreement
with the value derived by L\'opez-S\'anchez \& Esteban
(12900\,$\pm$\,700\,K). Our estimate of T$_e$([\OII]) 
is in agreement, although slightly lower, with the value derived by
L\'opez-S\'anchez \& Esteban (12000\,$\pm$\,500\,K) using the relation given
by \cite{1992AJ....103.1330G}. }

Recent studies in this galaxy suggest that the ionizing star cluster in knot B
would have the youngest population of the galaxy ($\sim$\,100\,Myr). This knot
presents a blue color and high UV emission, indicating a recent star-formation
activity, supported by the presence of Wolf-Rayet features
in its spectrum \citep{2010A&A...516A.104L}. This is in agreement with our
measurements of the equivalent width of the H$\beta$ emission line
[EW(H$\beta$)] (see Tables \ref{ratiostot 1}-\ref{ratiostot 5}), which 
suggest that knot B is the youngest one since it has the highest EW(H$\beta$)
\citep{2004MNRAS.348.1191T}. 

{On the other hand, we can assume from the analysis performed that the total
oxygen abundances derived for knots A and B are very similar. 
The ionic oxygen abundances of O$^{+}$/H$^+$ derived for knot B using the
long-slit spectrum and the global measure of the echelle data are in good
agreement taking into account the errors involved.
Under the same considerations, we can assume that the differences
between the O$^{2+}$/H$^+$ ionic abundances and the total oxygen abundances
derived for the long-slit and for the global measure of the echelle spectrum
of knot B are negligible. 
The total oxygen abundances derived from both instruments and from different
components for knots A, B, C, and E are in good agreement taking into account
the observational errors. }

The differences found between these star-forming knots
of \Haro are similar (or even smaller) to what is found in
other works which spatially resolve individual knots that belong to
\HII\ galaxies or 
Blue Compact Dwarf (BCD) galaxies \cite[see
  e.g.][]{Kehrig+08,Cairos+09b,Perez-Montero+09,Perez-Montero+11,Garcia-Benito+10,2011MNRAS.tmp..430H}. However,
in general, these differences were attributed to 
the observational uncertainties (pointing errors, seeing variations, etc.) or
errors associated to the reddening correction and flux calibration. The
oxygen abundance variations were therefore not assumed as statistically
significant, concluding that there is a possible common chemical evolution
scenario in all of them. There are even greater differences when comparing the
estimated abundances of the individual knots with those derived from the
integrated spectra of the galaxies. For instance, \cite{Cairos+09b} found for
the integrated spectrum of Mrk\,1418 a lower value of direct oxygen abundance
by about 0.35\,dex (equivalent to a factor of 2.2) than for knots 1 and 2 of
that galaxy. They pointed out that while this variation could reflect a
real abundance difference in different scales (kpc-sized aperture for the
integrated spectrum and sizes of the order of 100\,pc for individual \HII
regions), it may also be due to  
relatively large measurement uncertainties for the weak [\OIII] auroral
emission line. Fortunately, as in H11 (where we used a double beam
long-slit spectrograph), our data are not affected by pointing errors and
seeing variations, and the other observational uncertainties have a second
order effect, since the echelle spectrograph 
simultaneously acquire all the observed spectral range. Likewise, the errors 
associated with the measurements of the weak auroral emission lines are
relatively small, specially for [\OIII].



The estimated extinction for all the star-forming knots analysed using 
different instruments and kinematical components (including the global
measure) are very similar and consistent with low extinction values, found
in this kind of objects \cite[][and references therein]{2008PhDT........35H}.

\begin{figure*}
\centering
\includegraphics[trim=0cm 0cm 0cm 0cm,clip,angle=0,width=8cm,height=8cm]{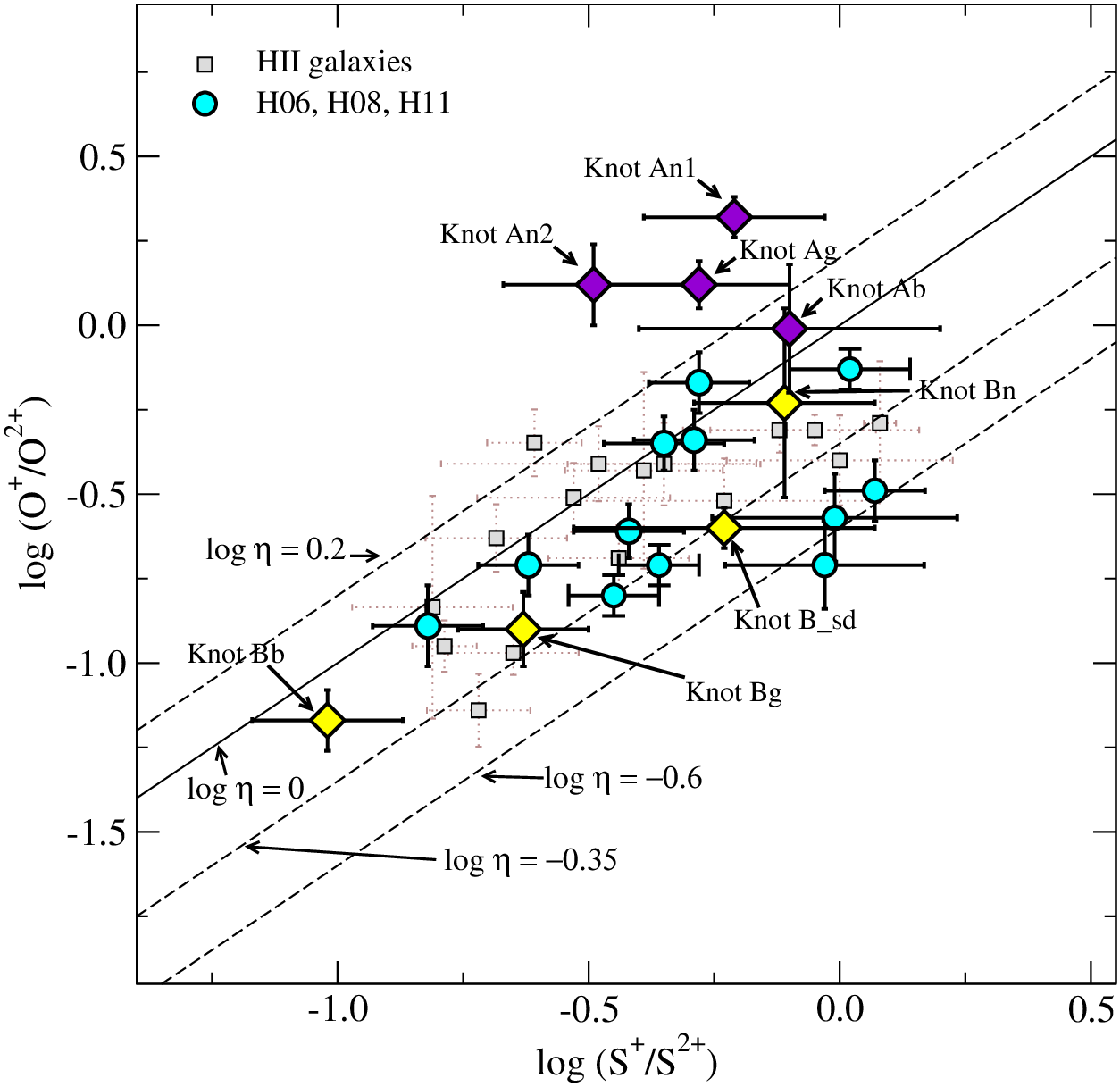}\hspace{0.3cm}
\includegraphics[trim=0cm 0cm 0cm 0cm,clip,angle=0,width=8cm,height=8cm]{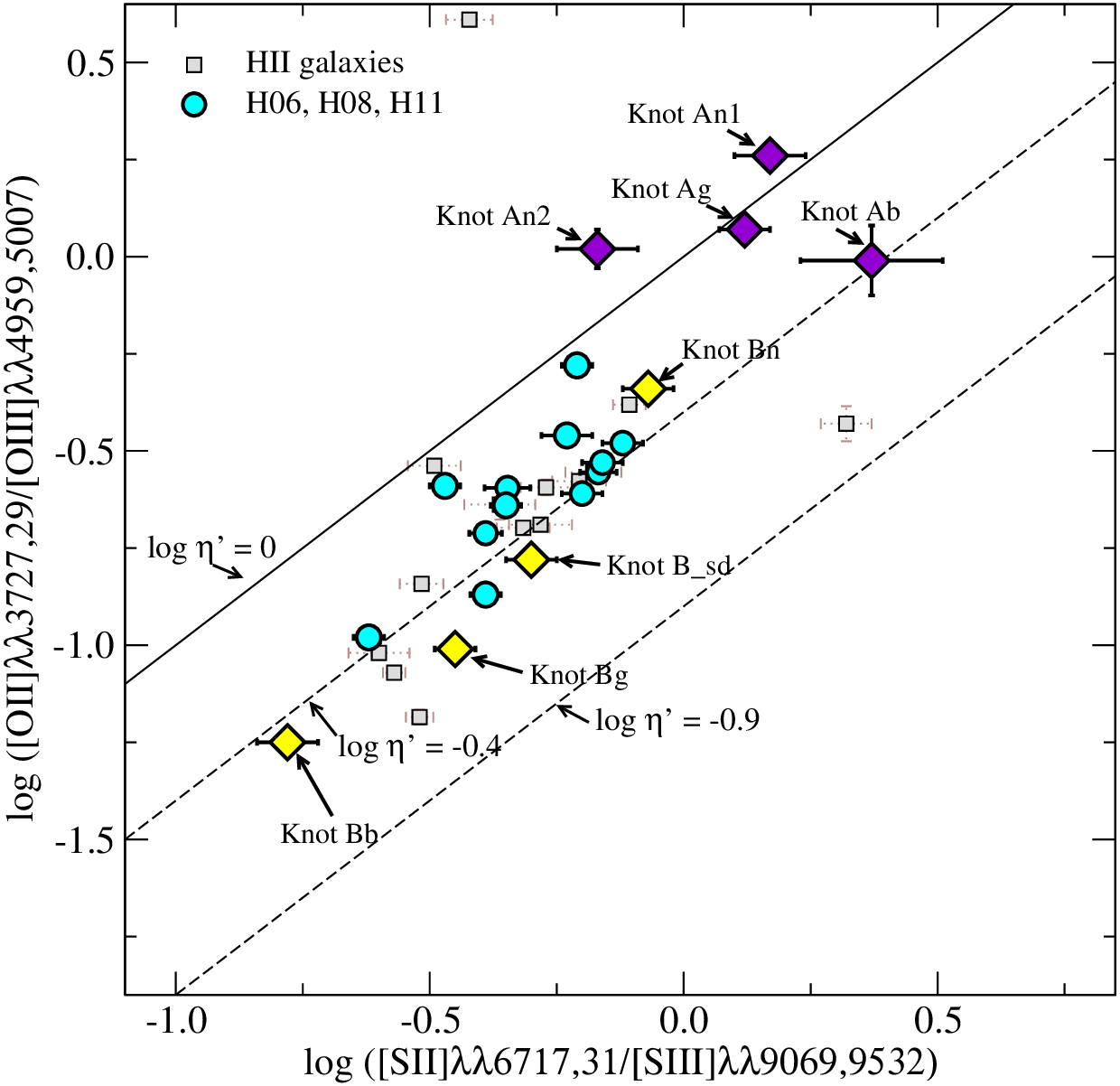}
\caption[]{Left panel: log(O$^+$/O$^{2+}$) vs.\ log(S$^{+}$/S$^{2+}$) for
  knots A and B (filled violet and yellow diamonds, respectively), the objects
  studied in H06, H08 and H11 (turquoise circles), and \HII galaxies from the
  literature as described in H08 (open squares). Diagonals in this diagram
  correspond to constant 
  values of $\eta$. Right panel: log([\OII]/[\OIII]) vs.\ log([\SII]/[\SIII]),
  symbols as in the left panel. Diagonals in this diagram correspond to
  constant values of $\eta$'. As in Fig.\ \ref{fig:abrelativas}, the letter
  added to the name of each knot denote the instrument (for long-slit
  spectrum) or the components (for echelle data).}
\label{fig:H15etaSO-ion}
\end{figure*}

\subsection{Ionization structure}
\label{ionization}

\cite{1988MNRAS.231..257V} showed that 
the quotient of O$^{+}$/O$^{2+}$ and S$^{+}$/S$^{2+}$, called ``softness
parameter" and denoted by $\eta$, is intrinsically related to the shape of the
ionizing continuum and depends only slightly on the geometry. 
We can perform an analysis of the ionization structure only for knots A and B
since we need information about two consecutive ionization stages of two
different atoms, reliably detected only in these two star-forming knots.

The purely observational counterpart of the $\eta$ parameter is the $\eta$'
parameter defined by \cite{1988MNRAS.231..257V} as the ratio between
[\OII]/[\OIII] and [\SII]/[\SIII]. The relation between both parameters is: 
\[ 
\log\,\eta'\,=\,\log\,\eta\,-\,0.14\,t_{e}\,-\,0.16
\]
where t$_{e}$ is the electron temperature in units of 10$^{4}$K. 

The left panel of Fig.\ \ref{fig:H15etaSO-ion} shows the relation between
log(O$^+$/O$^{2+}$) and log(S$^{+}$/S$^{2+}$) derived for knots A and B.
The global measures and the different kinematical components
of knot A and B are represented with violet and yellow diamonds,
respectively. In the right panel we show the relation between
log([\OII]/[\OIII]) vs.\ log\,([\SII]/[\SIII]), which does not require the 
explicit knowledge of the involved line temperatures for the derivation of
ionic ratios and is independent of the methods used  to estimate the line
temperatures. The objects studied by H06, H08, and H11 
are represented with turquoise circles, and the \HII galaxies from the
literature (see description and references in H08) with squares. The \HII
galaxies are located in the region corresponding to high values of the
effective temperatures of the radiation field of the ionizing star cluster,
where log\,$\eta$ are between -0.35 and 0.2 (see H06). In
Fig.\ \ref{fig:H15etaSO-ion} diagonal lines represent constant values  
of log\,$\eta$ (left panel) and log\,$\eta$'\, (right panel).

The ionization structures derived from both diagrams, $\eta$ and $\eta$', for
all the measurements (both instruments and different components) of knot B are
almost equal, showing almost the same values within the observational
errors. This implies that the effective temperatures of the radiation 
fields that ionize the gas  are very similar for different kinematical
components, then the ionizing star clusters that excite the gas that produce
the different kinematical components could be the same. For knot A we have
also derived very similar results from both diagrams. In this case
the broad kinematical component shows an ionization structure slightly
displaced in the $\eta$' diagram toward higher effective temperature of the ionizing radiation field
than those shown by both narrow kinematical components (and the global
measure). However, all the components are almost in the
same zone in the $\eta$ diagram if we take into account the large observational errors. This fact,
points again towards a common star cluster as the ionizing source for all the
components of knot A. The difference in the ionizing structure of these two
knots suggests a different evolutionary stage, since all the components of knot
A seem to be located in a region with lower effective temperature of the
ionizing radiation field than that where the components of knot B are placed. 
This is in agreement with the presence of an older and more evolved ionizing
star cluster in knot A than in knot B, as suggested by
\cite{2010A&A...516A.104L}.

\subsection{Chemical abundances from empirical calibrators}

Different strong line empirical metallicity calibrators are commonly used to
estimate the oxygen abundances in objects for which  direct derivation of
electron temperatures is not feasible. These empirical methods are based on
the cooling properties of ionized nebulae 
which ultimately translate into a relationship between emission line
intensities and oxygen abundance. These relationships have been widely studied
in the literature using different strong-line empirical methods which are
based on calibrations of the relative intensity of some bright emission lines 
against the abundance of relevant ions present in the nebula \cite[see
  e.g.][]{2008ApJ...677..201G,2009A&A...507.1291C,2009PhDT........15G,Garcia-Benito+10}. 

\begin{figure*}
\centering
\includegraphics[trim=0cm 0cm 0cm 0cm,clip,angle=0,width=0.8\textwidth]{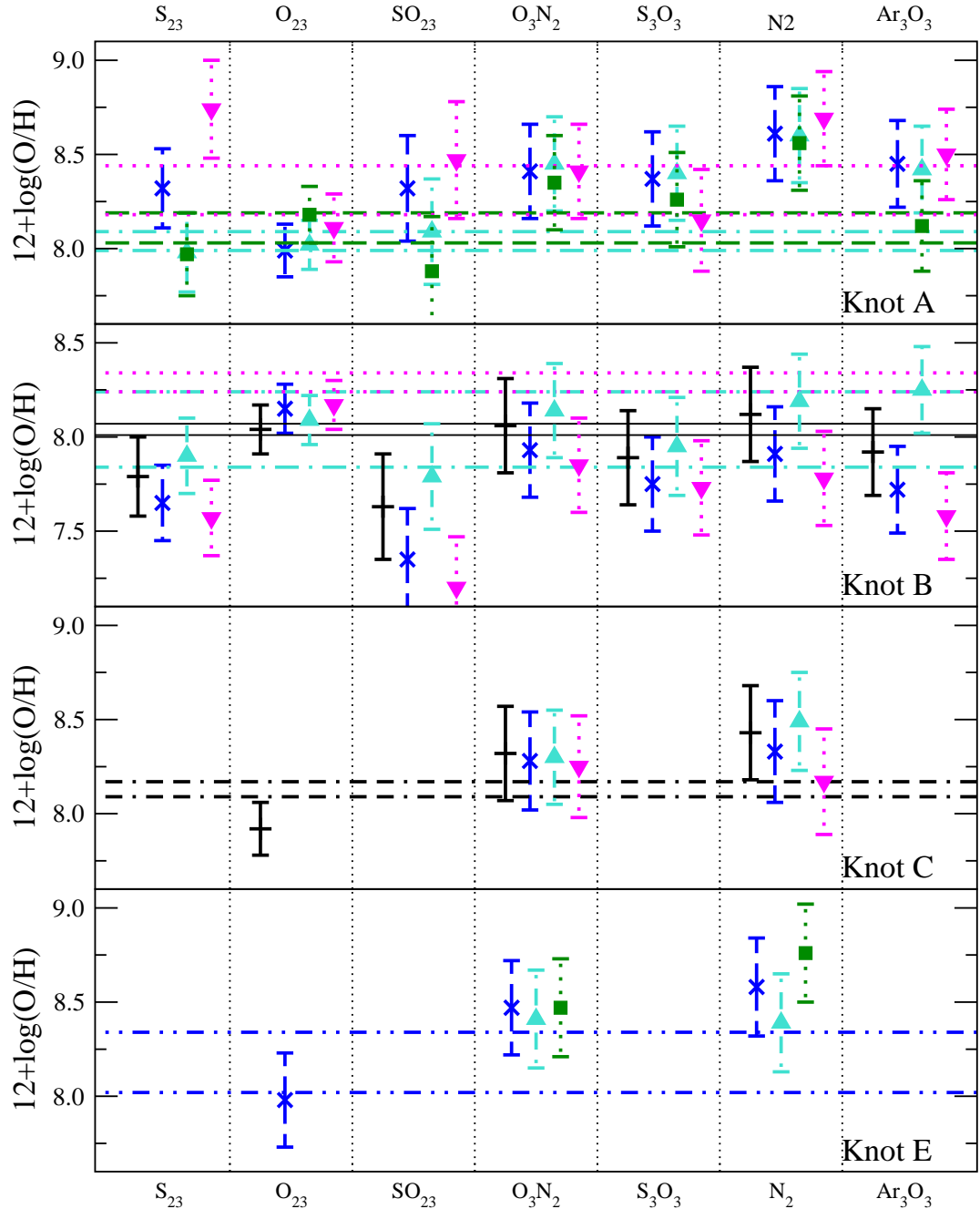}
\caption[]{The total oxygen abundances and their uncertainties for both
instruments, and the global measure and different kinematical components in
each observed knot of \Haro, as derived using different empirical
calibrators. From top to bottom panels: knots A, B, C, and E. From left to
right: the empirical parameters S$_{23}$, O$_{23}$, SO$_{23}$, O$_{3}$N$_{2}$,
S$_{3}$O$_{3}$, N$_{2}$, and Ar$_{3}$O$_{3}$. Solid black error bars represent
the long-slit estimate, blue stars: global echelle measure, turquoise up
triangles: narrow kinematical component (for knot A and E, filled up
triangles: narrow\,1 component, and filled green squares: narrow\,2 component),
and magenta down triangles: broad component. The solid lines represent the
valid range taking into account the 1\,$\sigma$ errors of the total oxygen
abundances previously derived using the direct method, and the non-continuous
lines the valid range of the total oxygen abundances derived using
photoionization models and/or empirical temperature relations (see text).}
\label{fig:calibradores empiricos}
\end{figure*}

{In Fig.\ \ref{fig:calibradores empiricos}, we show the total oxygen abundances
as derived from several strong-line empirical methods, with their
corresponding errors estimated taking into account the errors of the line
intensities and also the errors given by the calibrations of the empirical
parameters.
We have also plotted the total oxygen abundances calculated from the electron
temperatures measured using the direct method,
only for the long-slit data of knot B, and those estimated from the photoionization models
and empirical temperature relations.
The colours of these non-continuous lines (see the
{electron} edition of the journal) correspond to each measure or component,
thus the black solid and double-dashed-dotted lines represent the total oxygen
abundances calculated from the long-slit data, and turquoise dashed-dotted
(green dashed) and magenta dotted lines represent the narrow [narrow\,1
(turquoise) and narrow\,2 (green) for knot A and E] and broad components from
the echelle data, respectively. 
For simplicity, the abundance ranges corresponding to the global measure for
knot A and B are not shown.
This value is plotted only for knot E since it is the only
previous oxygen abundance estimation for this knot.}

{Among the available strong-line empirical parameters to obtain the total
oxygen abundance we studied the S$_{23}$ \citep{1996MNRAS.280..720V} using the 
calibration obtained by \cite{2005MNRAS.361.1063P}; 
O$_{23}$ \citep{Pagel+79}, and according to the values measured in our data,
we used the \cite{McGaugh91} calibration;
SO$_{23}$ \citep[\,=\,S$_{23}$/O$_{23}$;][]{2000MNRAS.312..130D} applying the
calibration from \cite{2005MNRAS.361.1063P}; 
O$_{3}$N$_{2}$ \citep{1979A&A....78..200A} with the calibration established by 
\cite{2004MNRAS.348L..59P};
S$_{3}$O$_{3}$ \citep{2006A&A...454L.127S};
N$_{2}$ \citep{Storchi-Bergmann+94} using the empirical calibration from
\cite{2002MNRAS.330...69D}; 
and Ar$_{3}$O$_{3}$ \citep{2006A&A...454L.127S}.}

{In the case of knot A, there is a good agreement between the total oxygen
abundance derived from the global measure using the different parameters and
the abundance derived using the temperatures obtained from models. Similar results
are found when using empirical temperature relations from the global measure.
For the narrow\,1 kinematical component there is a good agreement in the
abundance 
determination using the empirical parameters: S$_{23}$, O$_{23}$, and
SO$_{23}$, and that estimated using the temperature derived by models and/or
empirical temperature relations,
while the
derived abundance using the other empirical parameters
show small overestimations except the N$_{2}$ parameter which presents the
largest difference. 
The derived
abundance for the narrow\,2 component using the empirical parameters is in
agreement with that derived using the temperatures obtained from models and/or
empirical relations, except in the case of the N$_{2}$ parameter.
The derived abundances
from the broad component using different empirical parameters are very similar
to that derived using temperatures obtained from models or empirical relations.
The exception is the S$_{23}$ parameter that yields an abundance slightly
higher. 
}

{For the echelle data of knot B, the oxygen abundance calculated for the
global measure using the empirical parameters: O$_{23}$, O$_{3}$N$_{2}$, and
N$_{2}$ shows a similar value to that derived using the temperatures
obtained from models and/or empirical relations.
The estimated abundance from the other empirical parameters
present lower values than the abundance derived using the
line temperatures. 
The oxygen abundance for the narrow component derived using the
empirical parameters present similar values than that estimated using the line
temperatures,
and for the broad component the empirical parameters always present lower
values than the one determined using the derived temperatures, except for the
O$_{23}$ that shows a similar estimation.
For the single dispersion data, the total oxygen abundance derived from
empirical parameters are generally in good agreement whit that calculated
using the temperature obtained using the direct method. 
Only the abundance determined using the 
SO$_{23}$ empirical parameter is slightly lower than the previously estimated.
}

{For knot C, the total oxygen abundance derived using the empirical 
parameters 
are in good agreement with the oxygen
abundance value determined using the temperatures derived by models and/or
empirical temperatures relations for the long-slit data.
For knot E, there is a good agreement between the abundances derived from
the echelle measurements using the 
empirical parameters and that estimated using the adopted line temperatures, 
except for the N$_2$ parameter using the narrow\,2 component.
}


We must stress that the different empirical metallicity relationships were
derived from global measures of complete regions, and an analysis or
calibration for kinematical components has never been done before. To
disentangle the empirical relation for these two components, it is mandatory
to observe star-forming regions with high spectral resolution, similar to the
data presented in our work. Furthermore, even better S/N will be needed, in
particular to measure the auroral O$^+$ emission lines, to be able to obtain
direct measurements of the total oxygen abundances for the different
kinematical components. 
It must be noted that empirical relations for the estimation of abundances
should be applied only in a statistical sense and their application to
individual objects or regions is always questionable. 

\section{Summary and Conclusions}
\label{secConclus}
We have performed an analysis of the characteristics of the ionized gas
belonging to the four brightest star-forming knots of the BCD galaxy
\Haro. For two of these regions we present new low-resolution long-slit data
covering from 3800 to 9300\,\AA\ obtained with the Wide-Field CCD camera
mounted on the duPont Telescope at LCO. For all the regions we have carried
out a chemodynamics study using high resolution (R\,$\simeq$\,25000) data
acquired with the echelle spectrograph mounted at the same telescope. These
high resolution data were previously presented in a kinematical study in Paper
I. Using the direct method, and model based and empirical temperature
relations described in detail in H08, we have estimated the electron densities
and temperatures, ionic and total chemical abundances for different atomic
species. This analysis was done on the global measurements performed to the
emission lines of the spectra obtained with both instruments, and on the
different kinematical components deconvolved in the emission line profiles in
the high resolution data. 

{For all observed knots, and for all the global measurements and kinematical
components, the electron densities were found to be well below the critical
density for collisional de-excitation, as is well known to occur in the
star-forming processes belonging to \HII\ galaxies.}

{From the single dispersion spectra, in knot B we measured four line
temperatures: T$_e$([\OIII]), T$_e$([\OII]), T$_e$([\SIII]) and T$_e$([\SII])
reaching high precision, with rms fractional errors of the order of 2\%, 3\%,
10\% and 14\%, respectively. 
For knot C, from model based and empirical temperature
relations, we estimated the electron temperatures: T$_e$([\OIII]),
T$_e$([\OII]), T$_e$([\SIII]), T$_e$([\SII]) and T$_e$([\NII]), although in
these cases the quoted errors for the line temperatures derived using model
based relationships are only formal errors calculated from the measured line
intensity errors applying error propagation formula, without assigning any
error to the temperature calibration itself.}

{From the echelle spectra of knot B, 
we measured T$_e$([\OIII]) and T$_e$([\SIII]) applying
the direct method.} We have reached high precision from the global measure and
the broad component for the first temperature, with rms fractional errors of
the order of 5 and 4 per cent, respectively, and slightly worse, 15 per cent
for the narrow component. We have also obtained good precision for the estimates
of the [\SIII] temperature, 13, 18, and 14 per cent from the global measure,
the narrow and the broad components, respectively. 
{This is the first time that physical conditions are directly
estimated for kinematical components of \HII galaxies.}

{Using the estimated values for the electron densities and temperatures and
a careful and realistic treatment of observational errors, we have estimated
ionic and total abundances of O, S, N, Ne, Ar, Fe and He.
For the echelle data, we have also been able to carry out a chemodynamics
analysis applying the direct method to the kinematical component decomposition
of the emission line profiles.
This kind of analysis had never been done until now for this type of objects.
The obtained total abundances of O, S, N, Ne, and Ar are in the typical range
found for \HII\ galaxies. The total oxygen abundance derived for knot B is in
very good agreement with the value estimated by
\cite{2009A&A...508..615L}. The total oxygen abundances derived for all the
measurements of the echelle data of this knot are in good agreement. 
For knot A, the derived total oxygen abundance
from the global measure is lower ($\sim$\,0.2\,dex) than the value found by
\cite{2009A&A...508..615L}. However, taking into account all the possible
sources of uncertainties 
we find this difference negligible within the errors. Total oxygen abundances
derived for knots A, B, C, and E are very similar among themselves considering
the observational errors.
The N/O ratios derived from the different components of knot A and the global
measure of knot E are in very good agreement, and show an excess with respect
to the typical values found for \HII\ galaxies. Knot B seems to also show an
excess in the N/O ratio, but it is not as evident as in the case of knot A.
For knots A and B, the relative abundances of N/O, S/O, Ne/O, and Ar/O for the
different kinematical components are very similar, these could be evidence for
a common or very similar chemical evolution for the different kinematical
components of each knot. It could be also indicative that the different
kinematical components are different phases of the same gas.}


The ionization structure of knots A and B mapped through the use of the $\eta$
and $\eta'$ diagrams shows very similar values within the errors for the
different components of each of these regions. Only the broad component of
knot A seems to have slightly higher effective temperature (lower value of the
parameters) than the other components of this knots, showing similar values
to those of knot B. These similarities between the ionization structure of the
different kinematical components implies that the effective temperatures of
the ionizing radiation fields are very similar for all the different
kinematical components, in spite of some small differences in the ionization
state of different elements. The ionizing star clusters that excite the
gas belonging to each star-forming knot that produce the different kinematical
components could therefore be the same. The difference in the ionizing
structure of 
these two knots suggests a different evolutionary stage, with knot A located
in a region with lower effective temperature than that where knot B are placed. 
This is in agreement with the presence of an older and more evolved stellar
population in knot A than in knot B, as suggested by
\cite{2010A&A...516A.104L}. 

{We have also derived the total oxygen abundances using several strong-line
empirical parameters. In general, the estimated abundances are consistent with
the derived abundances using the temperatures calculated by the direct method
and/or by  model based and empirical temperature relationships. 
We have to emphasize that all the strong-line empirical parameters
were derived from measurements of the total intensity of the emission lines
from entire regions, and it has never been done an analysis or calibration for
kinematical components. Observations with high spectral resolution and
much better S/N, in particular to measure the auroral O$^+$ emission lines,
are needed to develop the empirical relations for the different kinematical
components.}

\section*{Acknowledgments}

We acknowledge fruitful discussions with Enrique P\'erez-Montero.
We thank very much an anonymous referee for a thorough reading of the
manuscript and for suggestions that greatly improved its clarity.
We are grateful to the director and staff of LCO for technical assistance and
warm hospitality. This research has made use of the NASA/IPAC Extragalactic
Database (NED) which is operated by the Jet Propulsion Laboratory, California
Institute of Technology, under contract with the National Aeronautics and
Space Administration.
Support from the Spanish \emph{Ministerio de Educaci\'on y Ciencia}
(AYA2007-67965-C03-03, AYA2010-21887-C04-03), and partial support from the
Comunidad de Madrid under grant S2009/ESP-1496 (ASTROMADRID) is acknowledged. 
VF and GB would like to thank the hospitality of the Astrophysics Group of the
UAM during the completion of this work.


\label{biblio} 
\bibliography{biblos} 
\bibliographystyle{mn2e} 

\end{document}